\documentclass[10pt,twocolumn,prb,aps,amssymb,amsmath,superscriptaddress]{revtex4}
\usepackage{color,graphicx,ulem}
\usepackage{bm}

\newcommand{\mbf}[1]{\mathbf{#1}}
\newcommand{\lb}{\left<}
\newcommand{\rb}{\right>}
\newcommand{\parl}{\parallel}
\newcommand{\unit}[1]{\mathbf{\hat{#1}}}
\newcommand{\pd}[2]{\frac{\partial#1}{\partial#2}}

\begin{document}

\title{Linearized model Fokker--Planck collision operators for gyrokinetic simulations.\\
 II. Numerical implementation and tests}

\author{M.\ Barnes}
\email{mabarnes@umd.edu}
\affiliation{
Department of Physics, IREAP and CSCAMM, University of Maryland, College Park, Maryland 20742-3511
}
\author{I.\ G.\ Abel}
\email{i.abel07@imperial.ac.uk}
\affiliation{
Plasma Physics Group, Blackett Laboratory, Imperial College, London SW7 2AZ, UK
}
\affiliation{
Euratom/UKAEA Fusion Association, Culham Science Centre, Abingdon OX14 3DB, UK
}
\author{W.\ Dorland}
\affiliation{
Department of Physics, IREAP and CSCAMM, University of Maryland, College Park, Maryland 20742-3511
}
\author{D. R. Ernst}
\affiliation{
Plasma Science and Fusion Center, Massachusetts Institute of Technology, 167 Albany Street, NW16-258, Cambridge, MA 02139
}
\author{G.\ W.\ Hammett}
\affiliation{
Princeton Plasma Physics Laboratory, Princeton University, P.O. Box 451, Princeton, New Jersey 08543
}
\author{P. Ricci}
\affiliation{
Centre de Recherches en Physique des Plasmas - \'{E}cole Polytechnique F\'{e}d\'{e}rale de Lausanne, Association EURATOM-Conf\'{e}d\'{e}ration Suisse, CH-1015 Lausanne, Switzerland
}
\author{B. N. Rogers}
\affiliation{
Department of Physics and Astronomy, Dartmouth College, Hanover, New Hampshire 03755
}
\author{A.\ A.\ Schekochihin}
\affiliation{
Plasma Physics Group, Blackett Laboratory, Imperial College, London SW7 2AZ, UK
}
\author{T.\ Tatsuno}
\affiliation{
Department of Physics, IREAP and CSCAMM, University of Maryland, College Park, Maryland 20742-3511
}

\date{\today}

\begin{abstract}

A set of key properties for an ideal dissipation scheme in gyrokinetic simulations is proposed,
and implementation of a model collision operator satisfying these properties is described.
This operator is based on the exact linearized test-particle collision operator, with 
approximations to the field-particle terms that preserve conservation laws and an H-Theorem.
It includes energy diffusion, pitch-angle scattering, and finite Larmor effects
corresponding to classical (real-space) diffusion.  The numerical implementation in the 
continuum gyrokinetic code GS2 is fully implicit and guarantees exact satisfaction of 
conservation properties.  Numerical results are presented showing that the correct physics is
captured over the entire range of collisionalities, from the collisionless to the strongly collisional regimes, without recourse to artificial 
dissipation.

\end{abstract}

\pacs{52.20.Hv,52.30.Gz,52.65.-y}

\maketitle

\section{Introduction}

Collisions play an important role in gyrokinetics.  An accurate collision operator is important
for calculation of neoclassical transport~\cite{boltonPoF83,xuPRE08} and the growth rate of 
instabilities such as trapped electron modes,~\cite{ernstPoP04,ernstIAEA06} dissipative
drift waves,~\cite{federiciPoF87,rewoldtPoF90,kotschCPC95} and microtearing modes~\cite{applegatePPCF07} in 
moderate collisionality regimes.
Collisions can also affect the damping of zonal flows~\cite{xiaoPoP07} and other modes
that provide a sink for turbulent energy.
In their absence, arbitrarily fine scales can develop in phase 
space,~\cite{krommesPop94,krommesPoP99,schekApJ08,schekPPCF08,barnesPoP08} 
which can in some cases pose challenges for discrete numerical algorithms, especially 
in the long-time limit;~\cite{grantPoF67,nevinsPoP05} even a modest amount of collisions
can make accurate numerical calculation much easier.  

Furthermore, inclusion of a small 
collisionality keeps the distribution function smooth enough in velocity space that the 
standard gyrokinetic ordering~\cite{friemanPoF82} for velocity space gradients is satisfied.
For example, the parallel nonlinearity,~\cite{villardFE04,candy2PoP06} given by
\begin{equation}
-\pd{}{v_{\parl}}\left[h\left(\frac{q}{m}\unit{b} + v_{\parl}\frac{\unit{b}\times \nabla B}{B^{2}}\right)\cdot \nabla \lb \varphi \rb\right]
\end{equation}
enters at the same order as the other terms in the gyrokinetic equation if 
the typical scale of parallel velocity fluctuations, $\delta v_{\parl}$, is one order 
smaller in the gyrokinetic expansion parameter $\rho/L$ ($\rho\equiv\textnormal{gyroradius}$ 
and $L\equiv\textnormal{background scale length}$) than the thermal speed, $v_{th}$.
Here, $h$ is the non-Boltzmann part of the perturbed distribution function (defined
more rigorously in the next section), $\varphi$ is
the electrostatic potential, $B$ is the magnetic field strength, $\unit{b}\equiv \mbf{B}/B$,
$q$ is particle charge, $m$ is particle mass, and $\lb \ . \ \rb$ denotes the gyroaverage at fixed
guiding center position $\mbf{R}$.

While such a situation is possible in the collisionless limit, a small collisionality prohibits
the formation of structures with $\delta v_{\parl}\sim (\rho/L)v_{th}$.  The level of
collisionality necessary to negate the importance of the parallel nonlinearity can be calculated
by assuming a balance between collisions and fluctation dynamics:
\begin{equation}
\pd{h}{t} \sim C[h] \Rightarrow \omega h \sim  \nu v_{th}^{2} \pd{^{2}h}{v^{2}},
\end{equation}
where $C[h]$ describes the effect of collisions on $h$, $\omega$ is the fluctuation frequency,
and $\nu$ is the collision frequency.  From the above expression, we see that scales 
in velocity space become small enough for the parallel nonlinearity to be important only
when the collision frequency satisfies $\nu \sim (\rho/L)^{2} \omega$.  Such
low collisionalities are not present in most fusion plasmas of interest.
Furthermore, if such an ordering had to be adopted, the lowest order distribution function 
could become strongly non-Maxwellian.  This is a problem for $\delta f$ 
codes that assume an equilibrium Maxwellian.


In light of the above considerations, it is important to include an accurate treatment of dissipation in gyrokinetic
simulations.  In order to faithfully represent gyrokinetic plasma dynamics at reasonable
numerical expense, we take the view that the form of the 
dissipation should be such that it: ensures satisfaction of the standard gyrokinetic ordering;
locally conserves particle number, momentum, and energy; satisfies Boltzmann's $H$-Theorem;
and efficiently smooths phase space structure.  The first of these requirements has already
been discussed in the context of the parallel nonlinearity.  Conservation properties have been 
found to be important, for instance, in calculations of the neoclassical ion thermal conductivity,~\cite{hintonRMP76}
as well as in a wide range of problems in fluid dynamics.  The existence of an $H$-Theorem
is critical for entropy balance~\cite{krommesPop94,sugamaPoP96,howesApJ06} and for the dynamics of the turbulent phase space cascade.~\cite{schekApJ08,schekPPCF08}
Efficient smoothing of phase space structures is necessary to resolve numerical simulations
at reasonable computational expense.  

A commonly employed dissipation mechanism in gyrokinetic simulations is
artificial (hyper) dissipation, often in physical (position) space.~\cite{kotschCPC95,jenkoPoP00,candyPoP06,barnesPoP08,chenPoP07}
Ideally, the form of the artificial dissipation should be chosen to satisfy the requirements
listed above and should be tested for convergence to the collisionless result.  Of course,
artificial dissipation alone is unable to capture the correct dynamics for moderate to strongly
collisional systems where turbulent fluxes and other observable quantities depend
sensitively on collisionality; for such systems, a physical dissipation model is desired.

A number of such model physical collision operators are employed in gyrokinetic codes.~\cite{kotschCPC95,jenkoPoP00,candyPoP06,navkalAPS06}
These range in complexity from the Krook operator~\cite{rewoldtPoF86} to the 
Rutherford-Kovrizhnikh operator~\cite{rutherfordPRL70} to the Catto-Tsang operator,~\cite{cattoPoF76} 
all of which have previously been implemented in GS2 (see, e.g. Ref.~\onlinecite{navkalAPS06}).
However, none of these satisfy all of the properties we require of a good collision operator (See
Appendix C of the accompanying paper~\cite{abelPoP08} for a fuller discussion of this point).
Here, we discuss numerical implementation in GS2 of an improved model 
operator which: includes the effects of both pitch-angle scattering and energy diffusion 
(i.e. efficiently smooths in phase space and ensures gyrokinetic ordering); 
conserves particle number, momentum, and energy;
satisfies Boltzmann's $H$-Theorem; and reduces to the linearized Landau test-particle operator 
in the large $k_{\perp}\rho$ limit.  A full description of this operator and a discussion
of its desirable properties, including comparison with the previous models of Catto-Tsang
and Hirshman-Sigmar,~\cite{hirshmanPoF76} is given in the companion paper~\cite{abelPoP08} 
(henceforth Paper I).  
We will focus on 
how such an operator can be implemented efficiently in gyrokinetic codes while maintaining
the properties listed above and on how our gyrokinetic dissipation scheme (or any other)
might be tested against a number of plasma physics problems.

The paper is organized as follows: in Section II, we introduce the gyroaveraged collision
operator from Paper I and examine properties
that should be taken into account when using it in numerical simulations;
in Section III, we describe our numerical implementation of the collision operator; in Section
IV we present numerical results for a number of tests demonstrating the ability of
our collision operator implementation to reproduce correct collisional and collisionless physics;
and in Section V, we summarize our findings.  


\section{Properties of the gyroaveraged collision operator}

In order to include collisions in gyrokinetics, we follow the treatment of Ref.~\onlinecite{howesApJ06} 
and assume the collision frequency, $\nu$, to be the same order 
in the gyrokinetic ordering as the characteristic fluctuation
frequency, $\omega$.\footnote{Note that this ordering does not prevent one from considering
the cases of $\nu \ll \omega$ and $\nu \gg \omega$ as subsidiary orderings.~\cite{schekApJ08}}
This leads to the requirement that the distribution of particles in
velocity space is Maxwellian to lowest order and allows us to represent the total distribution
function through first order in $\rho/L$ (where $\rho$ is ion gyroradius and $L$ is the scale
length of equilibrium quantities) as
\begin{equation}
f(\mbf{r},\mu,\varepsilon,t) = F_{0}(\varepsilon)\left(1-\frac{q\varphi(\mbf{r},t)}{T_{0}}\right) + h(\mbf{R},\mu,\varepsilon,t),
\end{equation}
where $\mbf{r}$ is particle position, $\mbf{R}=\mbf{r}-\unit{b}\times\mbf{v}/\Omega_{0}$ is guiding center position, 
$\mu\equiv mv_{\perp}^{2}/2B_{0}$ is magnetic moment, 
$\varepsilon\equiv mv^{2}/2$ is particle energy, $F_{0}$ is a Maxwellian,
$\varphi$ is the electrostatic potential, $B_{0}$ is the magnitude of the background 
magnetic field, $T_{0}$ is the background temperature, $q$ is particle charge, and
$\Omega_{0}=qB_{0}/mc$.  
The gyrokinetic equation governing the evolution of $h$ is given by
\begin{equation}
\begin{split}
\pd{h}{t}& + \left(v_{\parl}\unit{b} + \mbf{v}_{D}\right)\cdot \pd{h}{\mbf{R}} 
+ \frac{c}{B_{0}}\{\lb\chi\rb_{\mbf{R}},h\}\\
&= -q\pd{F_{0}}{\varepsilon}\pd{\lb\chi\rb_{\mbf{R}}}{t}
+\frac{c}{B_{0}}\{F_{0},\lb\chi\rb_{\mbf{R}} \} + \lb C[h] \rb_{\mbf{R}},
\end{split}
\end{equation}
where $\unit{b}\equiv \mbf{B}_{0}/B_{0}$, $\mbf{v}_{D}$ is the drift velocity of
guiding centers, $\chi\equiv \varphi - \mbf{v}\cdot\mbf{A}/c$, $\mbf{A}$ is the 
vector potential, $\{a,b\}$ is the Poisson bracket of $a$ and $b$, $\lb a \rb_{\mbf{R}}$ is 
the gyroaverage of $a$ at constant $\mbf{R}$, and $\lb C[h] \rb_{\mbf{R}}$ is the 
gyroaveraged collision operator.  

For $\lb C[h] \rb_{\mbf{R}}$, we restrict our attention to the model collision operator 
presented in Paper I.
We work within the framework of the continuum gyrokinetic code GS2,~\cite{kotschCPC95} 
which assumes periodicity in the spatial 
directions perpendicular to $\mbf{B}_{0}$ in order to reduce the simulation volume to a
thin flux tube encompassing a single magnetic field line.  Consequently, we require a spectral 
representation of $\lb C[h] \rb_{\mbf{R}}$:
\begin{equation}
\lb C[h] \rb_{\mbf{R}} \equiv \sum_{\mbf{k}} e^{i\mbf{k}\cdot\mbf{R}}C_{GK}[h_{\mbf{k}}],
\end{equation}
where $\mbf{k}$ is the perpendicular wavevector.
For convenience, we reproduce the expression for the same-species part of $C_{GK}[h_{\mbf{k}}]$ 
from Paper I in operator form:
\begin{equation}
C_{GK}[h_{\mbf{k}}] \equiv L[h_{\mbf{k}}] + D[h_{\mbf{k}}]
+ U_{L}[h_{\mbf{k}}] + U_{D}[h_{\mbf{k}}] + E[h_{\mbf{k}}],
\label{eqn:cop}
\end{equation}
where
\begin{equation}
L[h_{\mbf{k}}] \equiv \frac{\nu_{D}}{2}\pd{}{\xi}\left(1-\xi^{2}\right)\pd{h_{\mbf{k}}}{\xi} - \frac{k^{2}v^{2}}{4\Omega_{0}^{2}}\nu_{D}\left(1+\xi^{2}\right)h_{\mbf{k}}
\end{equation}
and
\begin{equation}
D[h_{\mbf{k}}] \equiv \frac{1}{2v^{2}}\pd{}{v}\left(\nu_{\parl}v^{4}F_{0}\pd{}{v}\frac{h_{\mbf{k}}}{F_{0}}\right) - \frac{k^{2}v^{2}}{4\Omega_{0}^{2}}\nu_{\parl}\left(1-\xi^{2}\right)h_{\mbf{k}}
\end{equation}
are the gyroaveraged Lorentz and energy diffusion operators (which together form the 
test-particle piece of the linearized Landau operator, as shown in Refs~\onlinecite{cattoPoF76}
and~\onlinecite{abelPoP08}),
\begin{equation}
\begin{split}
U_{L}[h_{\mbf{k}}] \equiv \nu_{D}F_{0}\Big(&J_{0}(\alpha)v_{\parl}
\frac{\int d^{3}v \ \nu_{D}v_{\parl}J_{0}(\alpha)h_{\mbf{k}}}{\int d^{3}v \ \nu_{D}v_{\parl}^{2}F_{0}} \\
&+ J_{1}(\alpha)v_{\perp}\frac{\int d^{3}v \ \nu_{D}v_{\perp}J_{1}(\alpha)h_{\mbf{k}}}{\int d^{3}v \ \nu_{D}v_{\parl}^{2}F_{0}}\Big)
\label{eqn:scriptuL}
\end{split}
\end{equation}
and
\begin{equation}
\begin{split}
U_{D}[h_{\mbf{k}}] \equiv -\Delta \nu &F_{0}\Big(J_{0}(\alpha)v_{\parl}
\frac{\int d^{3}v \ \Delta \nu v_{\parl}J_{0}(\alpha)h_{\mbf{k}}}{\int d^{3}v \ \Delta \nu v_{\parl}^{2}F_{0}} \\
&+ J_{1}(\alpha)v_{\perp}\frac{\int d^{3}v \ \Delta \nu v_{\perp}J_{1}(\alpha)h_{\mbf{k}}}{\int d^{3}v \ \Delta \nu v_{\parl}^{2}F_{0}}\Big)
\label{eqn:scriptuD}
\end{split}
\end{equation}
are the gyroaveraged momentum-conserving corrections to the Lorentz and energy diffusion
operators, and
\begin{equation}
E[h_{\mbf{k}}] \equiv \nu_{E} v^{2} J_{0}(\alpha) F_{0} \frac{\int d^{3}v \ \nu_{E} v^{2} J_{0}(\alpha) h_{\mbf{k}}}{\int d^{3}v \ \nu_{E} v^{4} F_{0}}
\label{eqn:scripte}
\end{equation}
is the gyroaveraged energy-conserving correction (the conserving terms are an approximation 
to the field-particle piece of the linearized Landau operator).  The electron
collision operator has the following additional term to account for electron-ion collisions:
\begin{equation}
\begin{split}
C_{GK}^{ei}[h_{e,\mbf{k}}] = \nu_{D}^{ei} \Big(\frac{1}{2}\pd{}{\xi}&\left(1-\xi^{2}\right)\pd{h_{e,\mbf{k}}}{\xi}-\frac{k^{2}v^{2}}{4\Omega_{0_{e}}^{2}}\left(1+\xi^{2}\right)h_{e,\mbf{k}}\\
&+\frac{2v_{\parl}u_{\parl}[h_{i,\mbf{k}}]}{v_{th_{e}}^{2}}J_{0}(\alpha_{e})F_{0e}\Big),
\label{eqn:Cei}
\end{split}
\end{equation}
where $\xi\equiv v_{\parl} / v$ is the pitch angle, $\alpha\equiv kv_{\perp}/\Omega_{0}$, $J_{0}$ and $J_{1}$
are Bessel functions of the first kind, $v_{th}\equiv\sqrt{2T_{0}/m}$ 
is the thermal velocity, and $u_{\parl}[h_{i,\mbf{k}}]$ is the perturbed parallel ion flow 
velocity.  Expressions for the velocity-dependent collision frequencies $\nu_{D}$, $\Delta \nu$, 
$\nu_{\parl}$, and $\nu_{E}$ are given in Paper I (which follows the notation of
Ref.~\onlinecite{hirshmanPoF76}).

Having specified the form of our collision operator, we now discuss some of its
fundamental properties that guide our choice of numerical implementation.

\subsection{Collision operator amplitude}

Even when the collisionality approaches zero, $C_{GK}[h_{\mbf{k}}]$ can have 
appreciable amplitude.  There are two reasons for this: first, the velocity dependence
of $\nu_{D}$, $\nu_{E}$, and $\Delta \nu$ is such that each go to infinity 
as $v\rightarrow0$ (so low-velocity particles are always collisional); and second,
we expect the distribution function to develop increasingly smaller scales in $v$ and $\xi$ as 
collisionality decreases, so that 
the amplitude of the terms proportional to $\partial^{2}h/\partial\xi^{2}$ and $\partial^{2}h/\partial v^{2}$ may
remain approximately constant\footnote{This is analagous to the result in fluid turbulence 
where the dissipation rate remains finite as viscosity becomes vanishingly small.} 
(i.e. $C_{GK}[h_{\mbf{k}}] \not\rightarrow 0$ as $\nu \rightarrow 0$).~\cite{krommesPop94, krommesPoP99, schekApJ08,barnesPoP08}  
The fact that $C_{GK}[h_{\mbf{k}}]$ can be quite large even at very low collisionalities
means that it should be treated implicitly if one wants to avoid a stability limit on the
size of the time step, $\Delta t$.  In Section III, we describe our fully implicit 
implementation of the collision operator.

\subsection{Local moment conservation}

Since collisions locally conserve particle density, momentum,
and energy, one would like these properties to be guaranteed by the discrete version
of the collision operator.  Mathematically, this means that the density, momentum, and energy moments
of the original (un-gyroaveraged) collision operator must vanish (for same-species
collisions).  However, the non-local nature of the gyroaveraging operation introduces finite 
Larmor radius (FLR) effects that lead to nonzero values for the analogous moments of 
$\lb C[h] \rb_{\mbf{R}}$.  Since this is the quantity we employ
in gyrokinetics, we need to find the pertinent relations its moments must satisfy in 
order to guarantee local conservation properties.

This is accomplished by Taylor expanding the Bessel functions $J_{0}$ and $J_{1}$.
In particular, one can show that~\cite{abelPoP08}
\begin{equation}
\begin{split}
&\int d^{3}v 
\begin{pmatrix}
1 \\
\mbf{v}\\
v^{2}
\end{pmatrix} 
\lb\lb C[h] \rb_{\mbf{R}}\rb_{\mbf{r}}\\
&=\sum_{\mbf{k}}e^{i\mbf{k}\cdot\mbf{r}} \int d^{3}v
\begin{pmatrix}
1 \\
v_{\parl}\unit{b}\\
v^{2}
\end{pmatrix}
C_{GK}^{0}[h_{\mbf{k}}] - \nabla \cdot \Gamma_{C},
\end{split}
\end{equation}
where $\lb . \rb_{\mbf{r}}$ denotes a gyroaverage at fixed $\mbf{r}$,
$C_{GK}^{0}[h_{\mbf{k}}]$ is the operator of Eqn.~(\ref{eqn:cop}) with $k\rho=0$
(neglecting FLR terms but retaining nonzero subscripts $\mbf{k}$ for $h$), and $\Gamma_{C}$ 
is the collisional flux of number, momentum, and energy
arising from FLR terms.  Consequently, the density, momentum,
and energy moments of the gyrokinetic equation can be written in the conservative form
\begin{equation}
\pd{\mathcal{M}}{t} + \nabla \cdot \Gamma_{\mathcal{M}} = \int d^{3}v  
\begin{pmatrix}
1 \\
v_{\parl}\unit{b}\\
v^{2}
\end{pmatrix} 
C_{GK}^{0}[h_{\mbf{k}}],
\end{equation}
where $\mathcal{M}\equiv \left(\delta n \ n\delta \mbf{u}_{\parl} \ \delta p\right)^{T}$
represents the perturbed number, momentum, and energy densities, the
superscript $T$ denotes the transpose, and $\Gamma_{\mathcal{M}}$ contains
both the collisional flux, $\Gamma_{C}$, and the flux arising from all other
terms in the gyrokinetic equation (for a more detailed discussion, see Paper I).
Thus, local conservation properties are assured in gyrokinetics as long as the density,
momentum, and energy moments of $C_{GK}^{0}[h_{\mbf{k}}]$ vanish:
\begin{equation}
\int d^{3}v
\begin{pmatrix}
1 \\
v_{\parl}\\
v^{2}
\end{pmatrix}
C_{GK}^{0}[h_{\mbf{k}}]=0.
\label{eqn:momcons}
\end{equation}  
We describe how this is accomplished numerically in Section III.


\subsection{$H$-Theorem}

In contrast with local conservation properties, the statement of the $H$-Theorem
is unmodified by gyroaveraging the collision operator.  Defining the entropy as 
$S=-f \ln f$, Boltzmann's $H$-Theorem tells us
\begin{equation}
\pd{S}{t} = -\int \frac{d^{3}\mbf{r}}{V}\int d^{3}v \ \ln[f] C[f] \ge 0,
\end{equation}
where $V\equiv \int d^{3}\mbf{r}$ and the double integration spans phase space (the velocity integration is taken at constant 
particle position $\mbf{r}$).  
Expanding the distribution function as before, we find to lowest order in the gyrokinetic ordering
\begin{equation}
\int \frac{d^{3}\mbf{r}}{V}\int d^{3}v \ \frac{h}{F_{0}} C[h] \le 0.
\end{equation}
Changing variables from particle position $\mbf{r}$ to guiding center position $\mbf{R}$,
we obtain
\begin{equation}
\int d^{3}v \int \frac{d^{3}\mbf{R}}{V} \ \frac{h}{F_{0}}  \lb C[h] \rb_{\mbf{R}} \le 0,
\label{eqn:gkdSdt}
\end{equation}
where now the velocity integration is taken at constant $\mbf{R}$.  In this case, the non-locality
of the gyroaveraging operation leads to no modification of the $H$-Theorem because of the definition
of entropy as a phase-space averaged quantity (as opposed to local conservation properties,
which involve only velocity-space averages).  Therefore, one can easily diagnose entropy
generation and test numerical satisfaction of the $H$-Theorem in gyrokinetic simulations, as
we show in Sec. IV.


\section{Numerical implementation}

It is convenient for numerical purposes to separately treat collisional and collisionless
physics.  Thus, we begin by writing the gyrokinetic equation in the form
\begin{equation}
\pd{h_{\mbf{k}}}{t} = C_{GK}[h_{\mbf{k}}] + \mathcal{A}[h_{\mbf{k}}],
\end{equation}
where $\mathcal{A}[h_{\mbf{k}}]$
represents the rate of change of $h_{\mbf{k}}$ due to the collisionless physics.  
In order to separate these terms, we utilize Godunov dimensional 
splitting,~\cite{GodunovMS59} which is accurate to first order in the timestep 
$\Delta t$:
\begin{gather}
\frac{h^{*}_{\mbf{k}}-h^{n}_{\mbf{k}}}{\Delta t} 
= \mathcal{A}[h^{n}_{\mbf{k}},h^{*}_{\mbf{k}}] \label{eqn:hstar}\\
\frac{h^{n+1}_{\mbf{k}}-h^{*}_{\mbf{k}}}{\Delta t}
=C_{GK}[h^{n+1}_{\mbf{k}}], 
\label{eqn:dhdtC}
\end{gather}
where $n$ and $n+1$ are indices representing the current and future time steps,
and $h^{*}_{\mbf{k}}$ is defined by Eqn.~(\ref{eqn:hstar}) -- it is the 
result of advancing the collisionless part of the gyrokinetic equation.  With $h^{*}_{\mbf{k}}$
thus given, we restrict our attention to solving Eqn.~(\ref{eqn:dhdtC}).
For notational convenience, we suppress all further $\mbf{k}$ subscripts, as we will be
working exclusively in $k$-space.

As argued in Section II, we must treat the collision operator implicitly to avoid a stability 
limit on the size of $\Delta t$.
We use a first order accurate backward-difference scheme in time 
instead of a second order scheme (such as Crank-Nicholson~\cite{richtmyer})
because it is well known that the Crank-Nicholson scheme
introduces spurious behavior in solutions to diffusion equations
when taking large timesteps (and because Godunov splitting is only first order
accurate for multiple splittings, which will be introduced shortly).

With this choice, $h^{n+1}$ is given by
\begin{equation}
h^{n+1} = \left(1-\Delta t  C_{GK}\right)^{-1}h^{*}.
\end{equation}
In general, $C_{GK}$ is a dense matrix, with both energy and pitch-angle
indices.  Inversion of such a matrix, which is necessary to solve for $h^{n+1}$ in our
implicit scheme, is computationally expensive.  We avoid this by taking two additional
simplifying steps.  First we employ another application of the Godunov splitting technique,
which, combined with the choice of a $(\xi,v)$ grid in GS2,~\cite{barnesPoP08} allows us to consider energy 
and pitch-angle dependence separately:~\footnote{We note that an implicit treatment of the Catto-Tsang operator (including energy diffusion)
has independently been implemented in GS2 using the same splitting technique.~\cite{navkalAPS06}}
\begin{gather}
h^{**} = \left[1-\Delta t \left( 
L + U_L\right)\right]^{-1} h^{*}
\label{eqn:Lsplit}
\\
h^{n+1}=\left[1 - \Delta t \left(D
+ U_D+ E\right)\right]^{-1}h^{**},
\label{eqn:DMEsplit}
\end{gather} 
The $h^{*}$ and $h^{**}$ are vectors whose 
components are the values of $h$ at each of the $(\xi,v)$ grid points.  In Eqn.~(\ref{eqn:Lsplit})
we order the components so that
\begin{equation}
h \equiv \left(h_{11},h_{21},...,
h_{N1},h_{12},...,h_{NM}\right)^{T},
\end{equation}
where the first index represents pitch-angle, the second represents energy, and 
$N$ and $M$ are the number of pitch-angle and energy grid points, respectively.
This allows for a compact representation in pitch-angle.  When solving Eqn.~(\ref{eqn:DMEsplit}),
we reorder the components of $h$ so that
\begin{equation}
h \equiv \left(h_{11},h_{12},...,
h_{1N},h_{21},...,h_{NM}\right)^{T},
\end{equation}
allowing for a compact representation in energy.

\subsection{Conserving terms}

The matrices $1-\Delta t L$ and 
$1-\Delta t D$ are chosen
to be tridiagonal by employing three-point stencils for finite differencing in $\xi$ and $v$.  
This permits computationally
inexpensive matrix inversion.  However, the full matrices to be inverted include the
momentum- and energy-conserving operators, $U$ and $E$, which are dense
matrices.  We avoid direct inversion of these matrices by employing the Sherman-Morrison
formula,~\cite{shermanAMS49,shermanAMS50} which gives $\mbf{x}$ in the matrix 
equation $M\mbf{x}=\mbf{b}$,
as long as $M$ can be written in the following form:
\begin{equation}
M = A + \mathbf{u}\otimes\mathbf{v},
\end{equation} 
where $\otimes$ is the tensor product.  The solution is then given by
\begin{equation}
\mbf{x} = \mbf{y} - \left[\frac{\mbf{v}\cdot\mbf{y}}{1+\mbf{v}\cdot\mbf{z}}\right]\mbf{z},
\end{equation}
where $\mbf{y}=A^{-1}\mbf{b}$, $\mbf{z}=A^{-1}\mbf{u}$,
and the dot products represent integrals over velocity space.  If $A^{-1}$ is known or
easily obtainable (as in our case), this formulation provides significant computational
savings over the straighforward method of directly inverting the dense matrix $M$.

Details of the application of the Sherman-Morrison formula to Eqns.~(\ref{eqn:Lsplit}) and
(\ref{eqn:DMEsplit}) are given in Appendix A.  Here, we state the main points.  
The matrix operators $L$ and $D$ are to be identified with $A$, and the integral conserving
terms $U$ and $E$ can be written in the form of the tensor product, $\mbf{u}\otimes\mbf{v}$. 
Identifying
$h^{**}$ and $h^{n+1}$ with $\mbf{x}$, we find that multiple applications of the 
Sherman-Morrison formula give
\begin{equation}
h = \mbf{y}_{2} - \left[\frac{\mbf{v}_{2}\cdot \mbf{y}_{2}}{1+\mbf{v}_{2}\cdot\mbf{z}_{2}}\right]\mbf{z}_{2},
\label{eqn:hx}
\end{equation}
where 
\begin{gather}
\mbf{y}_{2} = \mbf{y}_{0} - \left[\frac{\mbf{v}_{0}\cdot\mbf{y}_{0}}{1+\mbf{v}_{0}\cdot\mbf{s}_{0}}\right]\mbf{s}_{0} - \left[\frac{\mbf{v}_{1}\cdot\mbf{y}_{0}}{1+\mbf{v}_{1}\cdot\mbf{w}_{0}}\right]\mbf{w}_{0}, \label{eqn:sm1}\\
\mbf{z}_{2} = \mbf{z}_{0} - \left[\frac{\mbf{v}_{0}\cdot\mbf{z}_{0}}{1+\mbf{v}_{0}\cdot\mbf{s}_{0}}\right]\mbf{s}_{0}- \left[\frac{\mbf{v}_{1}\cdot\mbf{z}_{0}}{1+\mbf{v}_{1}\cdot\mbf{w}_{0}}\right]\mbf{w}_{0}. \label{eqn:sm2}
\end{gather}
The quantities $\mbf{v}_{0}$, $\mbf{v}_{1}$, $\mbf{v}_{2}$, $\mbf{z}_{0}$, $\mbf{s}_{0}$, $\mbf{w}_{0}$, and $\mbf{y}_{0}$
are specified in Table~\ref{tab:sm} in Appendix A.  With the exception
of $\mbf{y}_{0}$, each of these quantities
is time-independent, so they need be computed only once at the
beginning of each simulation.  Consequently, inclusion of the conserving terms in our
implicit scheme comes at little additional expense.

We note that when Eqn.~(\ref{eqn:hx}) is applied to computing the inverse matrix in 
Eqn.~(\ref{eqn:Lsplit}), the corresponding $\mbf{v}_{2}$ 
is nonzero only for the electron collision operator.
This term arises by using the parallel component of Ampere's law to rewrite the electron-ion
collison operator of Eqn.~(\ref{eqn:Cei}) as
\begin{equation}
\begin{split}
C_{GK}^{ei}[h_{e}] &= \nu_{D}^{ei}\Bigg(\frac{1}{2}\pd{}{\xi}\left(1-\xi^{2}\right)\pd{h_{e}}{\xi} - \frac{k^{2}v^{2}}{4\Omega_{0,e}^{2}}\left(1+\xi^{2}\right)h_{e}\\
&+\frac{2v_{\parl}}{v_{th_{e}}^{2}}J_{0}(\alpha_{e})F_{0e}\left[u_{\parl}[h_{e}]+\frac{ck^{2}}{4\pi e n_{0,e}}A_{\parl}\right]\Bigg),
\end{split}
\end{equation}
where $e$ is the magnitude of the electron charge, $u_{\parl}[h_{e}]$ is the parallel
component of the electron 
fluid velocity, and $n_{0,e}$ is the equilibrium electron density.
For electron collisions, the $A_{\parl}$ term is absorbed into $h^{*}$ so that we use
the modified quantity
\begin{equation}
\tilde{h}_{e}^{*} = h_{e}^{*} + \nu_{D}^{ei}\Delta t\frac{ck^{2}v_{\parl}}{2\pi e v_{th_{e}}^{2}n_{0,e}}A_{\parl}J_{0}(\alpha_{e})F_{0,e}
\end{equation}
when applying the Sherman-Morrison formula, where $A_{\parl}$ from the $n+1$ time
level is used (for details on the implicit calculation of $A_{\parl}$, see Ref.~\onlinecite{kotschCPC95}).

\subsection{Discretization in energy and pitch angle}

We still must specify our choice of discretization for $C_{GK}$.
Ideally, we would like the discrete scheme to guarantee the 
conservation properties and $H$-Theorem associated with $C$.  
As discussed in Sec. II, the former is equivalent to requiring that the $k\rho=0$ component
of $C_{GK}$, $C_{GK}^{0}$, satisfy Eqn.~(\ref{eqn:momcons}).  We now
proceed to show that this requirement is satisfied by carefully discretizing the conserving
terms and by employing a novel finite difference
scheme that incorporates the weights associated with our numerical integration scheme.

We begin by writing $C_{GK}^{0}[h]$ for same-species collisions:
\begin{widetext}
\begin{equation}
\begin{split}
C_{GK}^{0}[h]& = \frac{\nu_{D}}{2}\pd{}{\xi}\left(1-\xi^{2}\right)\pd{h}{\xi}
+\frac{1}{2v^{2}}\pd{}{v}\left(\nu_{\parl}v^{4}F_{0}\pd{}{v}\frac{h}{F_{0}}\right)
+ \nu_{D}v_{\parl}F_{0}\frac{\int d^{3}v \ \nu_{D} v_{\parl}h}{\int d^{3}v \ \nu_{D}v_{\parl}^{2}F_{0}}\\
&- \Delta\nu v_{\parl}F_{0}\frac{\int d^{3}v \ \Delta\nu v_{\parl}h}{\int d^{3}v \ \Delta\nu v_{\parl}^{2}F_{0}}
+ \nu_{E}v^{2}F_{0}\frac{\int d^{3}v \ \nu_{E}v^{2}h}{\int d^{3}v \ \nu_{E}v^{4}F_{0}}.
\label{eqn:cgk0}
\end{split}
\end{equation}
\end{widetext}
With $C_{GK}^{0}$ thus specified, we now consider numerical evaluation of the relevant moments of Eqn.~(\ref{eqn:cgk0}).  
To satisfy number conservation ($\int d^{3}v \ C_{GK}^{0}[h]=0$), velocity space
integrals of each of the terms in Eqn.~(\ref{eqn:cgk0}) should vanish individually.  For
integrals of the first two terms to vanish, we require a finite difference scheme that satisfies 
a discrete analog of
the Fundamental Theorem of Calculus (i.e. conservative differencing); for the last three
terms, we must have a discrete integration scheme satisfying 
$\int d^{3}v \ \nu_{D}v_{\parl}F_{0}=\int d^{3}v \ \Delta \nu v_{\parl}F_{0} 
= \int d^{3}v \ \nu_{E}v^{2}F_{0} = 0$. 
The requirement that $\int d^{3}v \ \nu_{D}v_{\parl}F_{0}=\int d^{3}v \ \Delta\nu v_{\parl}F_{0}=0$ 
is satisfied by any integration scheme with velocity space grid points and associated 
integration weights symmetric about $v_{\parl}=0$, which 
is true for the $(\xi,v)$ grid used in GS2.  By substituting for $\nu_{E}$ everywhere using
the identity
\begin{equation}
\nu_{E}v^{2}F_{0}=-\frac{1}{v^{2}}\pd{}{v}\left(\nu_{\parl}v^{5}F_{0}\right),
\label{eqn:nue1}
\end{equation}
the other integral constraint
($\int d^{3}v \ \nu_{E}v^{2}F_{0}=0$) reduces to the requirement that finite difference
schemes must satisfy the Fundamental Theorem of Calculus.

Parallel momentum conservation ($\int d^{3}v \ v_{\parl}C_{GK}^{0}[h]=0$) 
introduces the additional requirements that:
\begin{equation}
\int d^{3}v \ v_{\parl}\left(\frac{\nu_{D}}{2}\pd{}{\xi}\left(1-\xi^{2}\right)\pd{h}{\xi}
+ v_{\parl}\nu_{D}F_{0}\frac{\int d^{3}v \ \nu_{D}v_{\parl}h}{\int d^{3}v \ \nu_{D}v_{\parl}^{2}F_{0}}\right)=0
\label{eqn:vpaL}
\end{equation}
and
\begin{equation}
\begin{split}
\int d^{3}v \ &\frac{v_{\parl}}{2v^{2}}\pd{}{v}\left(\nu_{\parl}v^{4}F_{0}\pd{}{v}\frac{h}{F_{0}}\right)\\
&= \int d^{3}v \ \Delta\nu v_{\parl}^{2}F_{0}\frac{\int d^{3}v \ \Delta\nu v_{\parl}h}{\int d^{3}v \ \Delta\nu v_{\parl}^{2}F_{0}}.
\end{split}
\label{eqn:vpaD}
\end{equation}
If the finite difference scheme used for all differentiation possesses a discrete version of 
integration by parts (upon double 
application), then Eqns.~(\ref{eqn:vpaL}) and $(\ref{eqn:vpaD})$ are numerically 
satisfied as long as:
$v_{\parl}\nu_{D}h$ in the second term of Eqn.~(\ref{eqn:vpaL}) is expressed in the form
\begin{equation}
v_{\parl}\nu_{D}h = -\frac{1}{2}\left(\pd{}{\xi}\left(1-\xi^{2}\right)\pd{v_{\parl}}{\xi}\right)\nu_{D}h,
\label{eqn:vpa}
\end{equation}
$\Delta \nu$ on the righthand side of Eqn.~(\ref{eqn:vpaD}) is expressed using the identity
\begin{equation}
2\Delta \nu v^{3}F_{0} = \pd{}{v}\left(\nu_{\parl}v^{4}F_{0}\pd{v}{v}\right),
\label{eqn:delnu}
\end{equation}
and all integrals are computed using the same numerical integration scheme
(if analytic results for the integral denominators in terms 3 and 4 of Eqn.~(\ref{eqn:cgk0}) 
are used, then the necessary exact cancellation in Eqns.~(\ref{eqn:vpaL}) and (\ref{eqn:vpaD})
will not occur).

The only additional constraint imposed by energy conservation
($\int d^{3}v \ v^{2}C_{GK}^{0}[h]=0$) is that the form of Eqn.~(\ref{eqn:nue1})
be slightly modified so that
\begin{equation}
\nu_{E}v^{2}F_{0} = -\frac{1}{v^{2}}\pd{}{v}\left(\nu_{\parl}v^{4}F_{0}\pd{v^{2}}{v}\right),
\label{eqn:nue}
\end{equation}
which still satisfies the number conservation contraint.
Using the forms given by Eqns.~(\ref{eqn:vpa})-(\ref{eqn:nue}), conservation properties
are guaranteed as long as one
employs a finite difference scheme for pitch-angle scattering and energy diffusion that
satisfies discrete versions of the Fundamental Theorem of Calculus and integration by parts.

For the case of equally spaced grid points in $v$ and $\xi$, there is a straightforward 
difference scheme, accurate to secord order in the grid spacing, that satisfies both 
requirements:~\cite{degondNM94}
\begin{equation}
\pd{}{x}G\pd{h}{x} \approx \frac{G_{j+1/2}\left(h_{j+1}-h_{j}\right)-G_{j-1/2}\left(h_{j}-h_{j-1}\right)}{\Delta x^{2}},
\end{equation}
where $x$ is a dummy variable representing either $v$ or $\xi$, $\Delta x$ is the grid
spacing, $h_{j}$ is the value of $h$ evaluated at the grid point $x_{j}$, 
$x_{j\pm 1/2}\equiv (x_{j}+x_{j\pm 1})/2$, and $G$ is either $1-\xi^{2}$ (for 
pitch-angle scattering) or $\nu_{\parl}v^{4}F_{0}$ (for energy diffusion).
However, in order to achieve higer order accuracy in the calculation
of the velocity space integrals necessary to obtain electromagnetic fields,
GS2~\cite{barnesPoP08} and a number of other
gyrokinetic codes~\cite{candyJCP03} use grids with unequal spacing in $v$ and $\xi$ and integration weights
that are not equal to the grid spacings.  

Given the constraints
of a three-point stencil on an unequally spaced grid, we are forced to choose between a 
higher order scheme (a second order accurate scheme can be obtained with compact 
differencing,~\cite{durran} as described in Appendix B) that does not satisfy our two requirements and a lower order scheme that does.
Since our analytic expression for $C_{GK}$ was designed in large part to satisfy conservation 
properties (and because the conserving terms are only a zeroth order accurate approximation
to the field-particle piece of the linearized Landau operator~\cite{hirshmanPoF76}), we 
choose the lower order scheme, given here, as the default:
\begin{equation}
\pd{}{x}G\pd{h}{x} \approx \frac{1}{w_{j}}\left(G_{j+1/2}\frac{h_{j+1}-h_{j}}{x_{j+1}-x_{j}}-G_{j-1/2}\frac{h_{j}-h_{j-1}}{x_{j}-x_{j-1}}\right),
\label{eqn:fd}
\end{equation}
where $w_{j}$ is the integration weight associated with $x_{j}$.  

\begin{figure}
\includegraphics[height=1.65in]{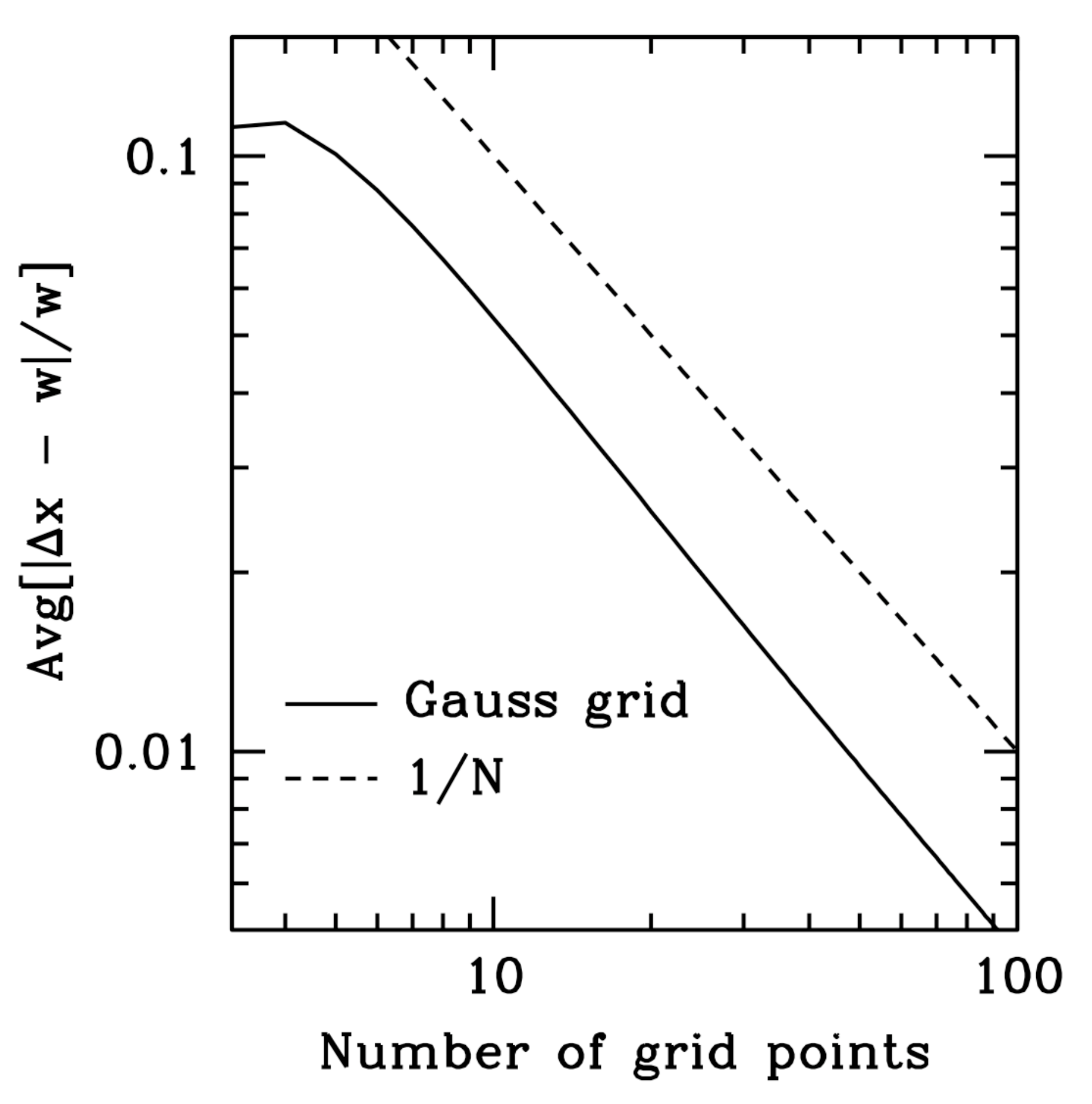}
\includegraphics[height=1.65in]{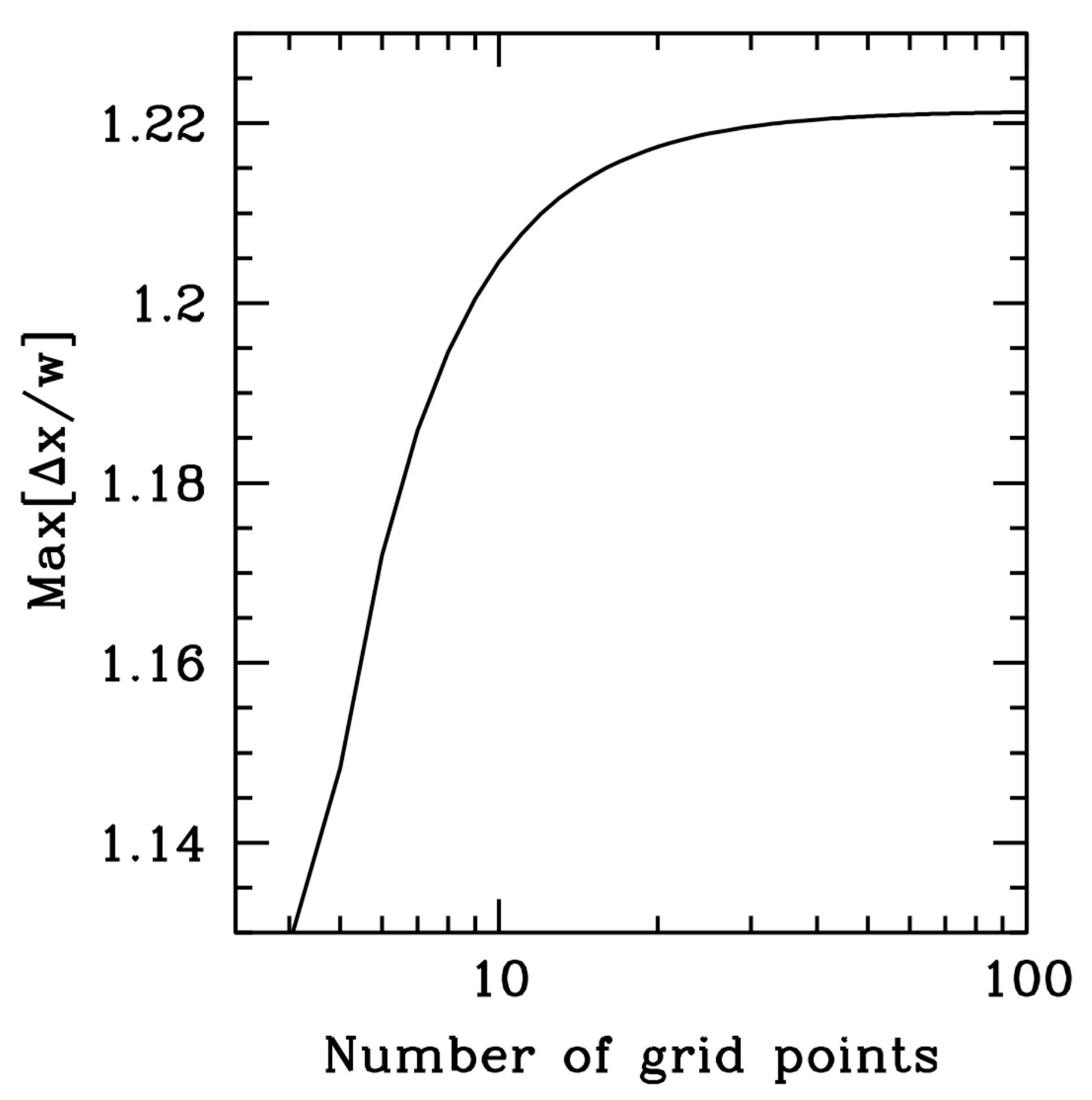}
\caption{(Left): Solid line indicates the scaling of the leading order error, averaged over 
all grid points, of the conservative finite difference scheme for a Gauss-Legendre grid (the 
grid used in GS2).  The slope of the dotted line corresponds to a first order scheme.
(Right): factor by which the conservative finite difference scheme
of Eqn.~(\ref{eqn:fd}) amplifies the true collision operator amplitude at the boundaries of 
the Gauss-Legendre grid.}
\label{fig:errfd}
\end{figure}

\begin{figure*} 
\begin{center}
\includegraphics[height=2.4in]{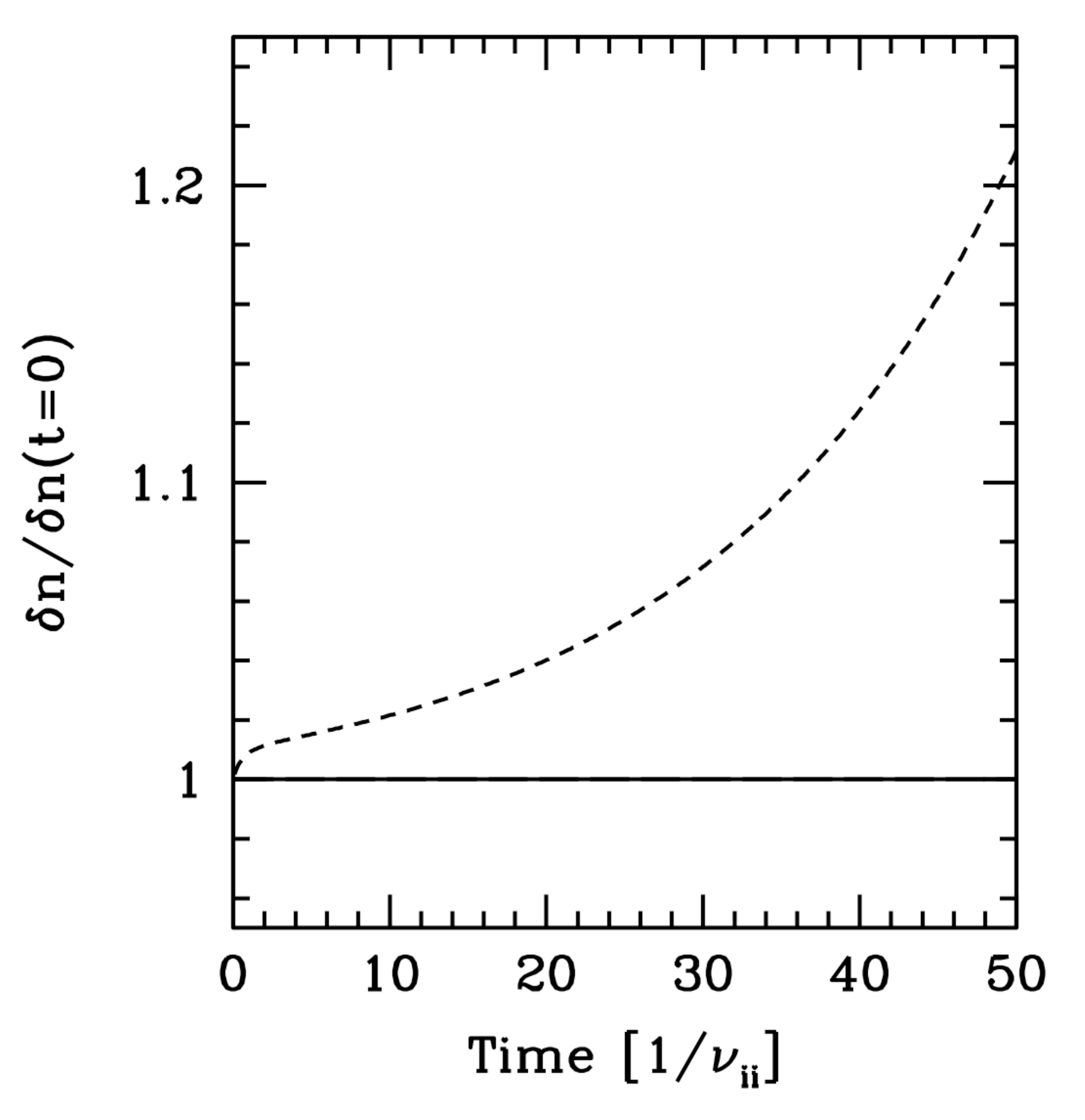}
\includegraphics[height=2.4in]{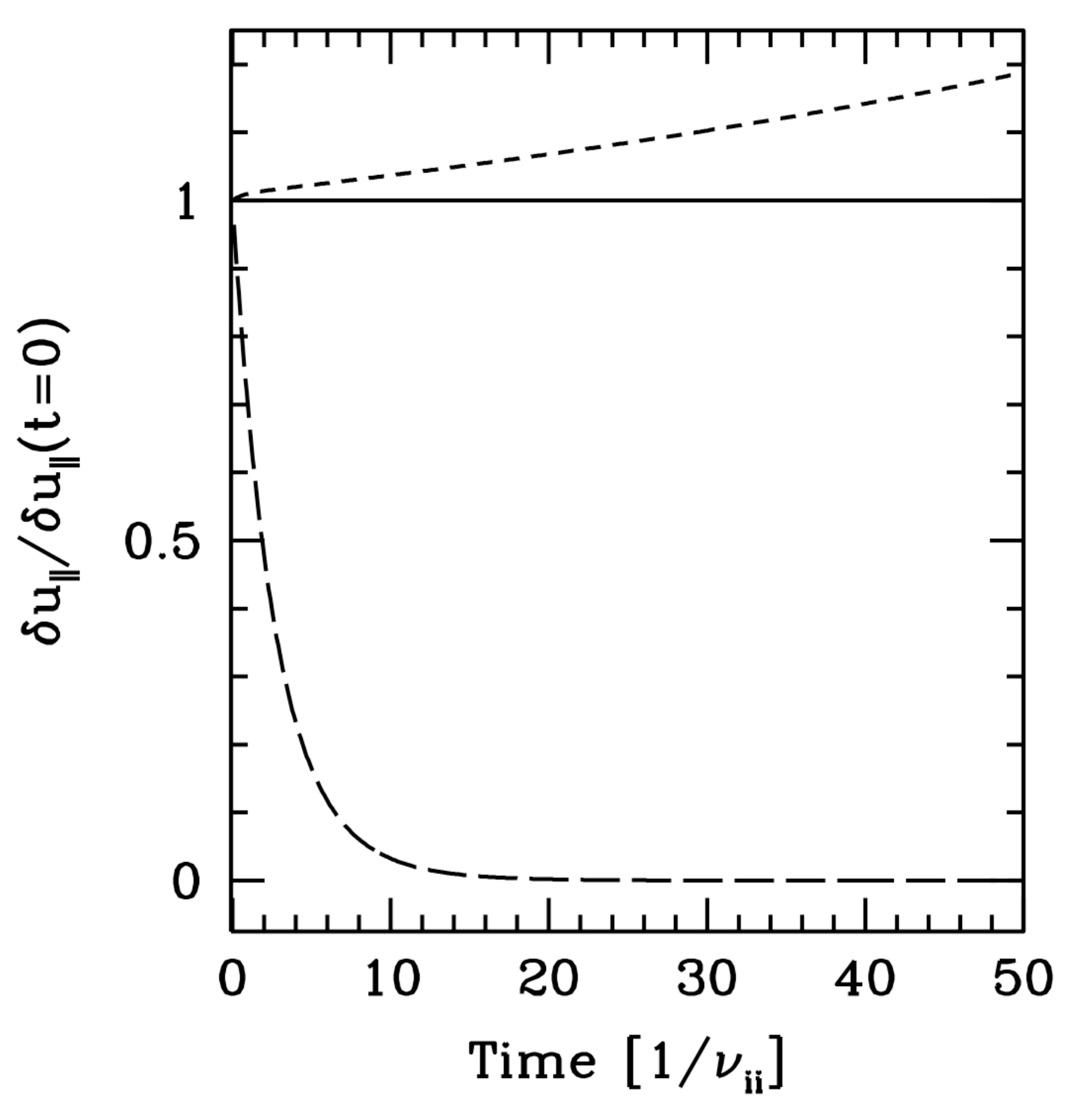}
\includegraphics[height=2.4in]{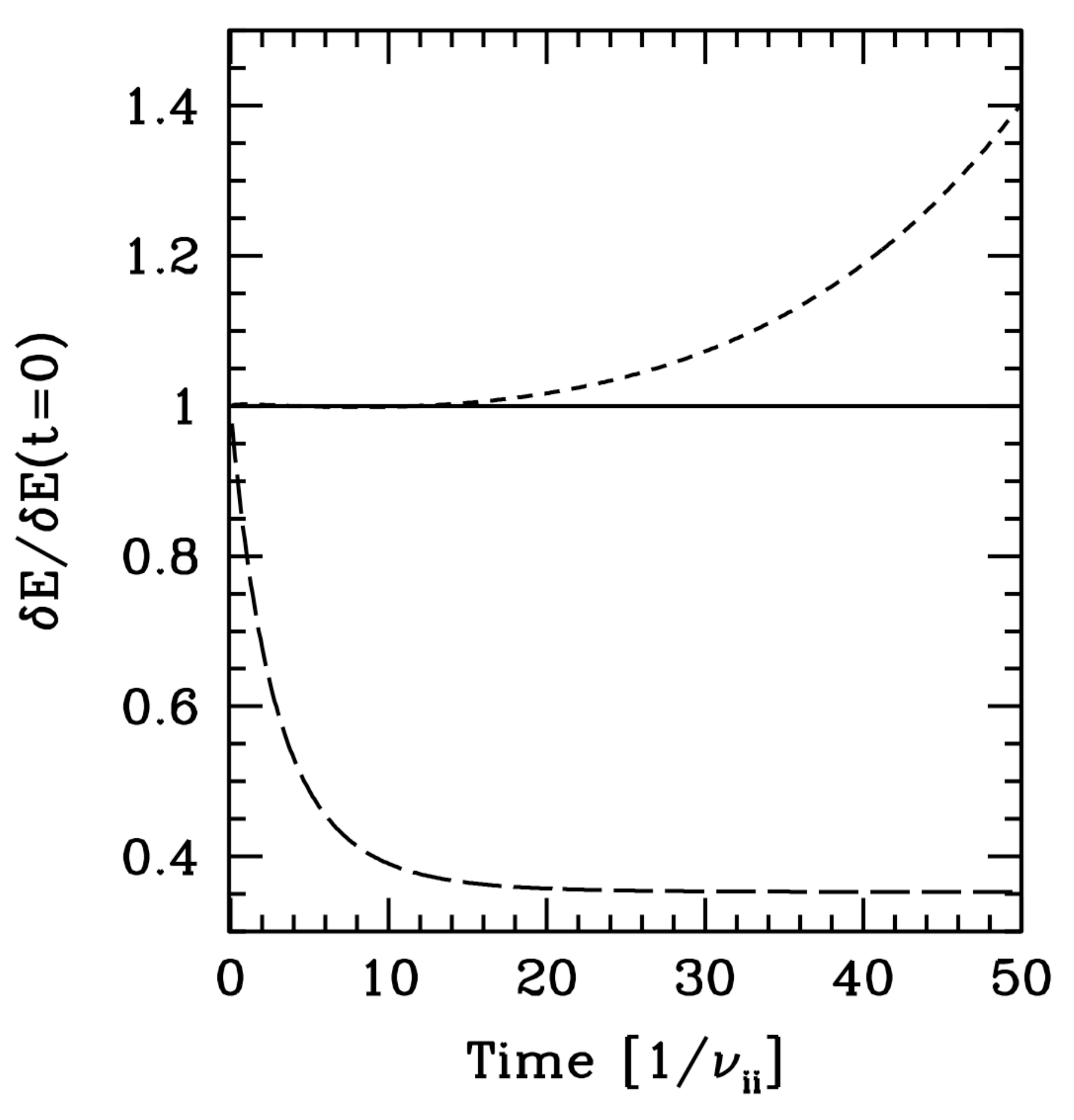}
\end{center}
\caption{Plots showing evolution of the perturbed local density, parallel momentum, and energy 
over fifty collision times.  Without the conserving terms (\ref{eqn:scriptuL})-(\ref{eqn:scripte}), 
both parallel momentum and energy decay significantly over a few collision times (long dashed lines).  
Inclusion of conserving terms with the conservative scheme detailed in Sec. III leads to exact 
moment conservation (solid lines).  Use of a non-conservative scheme leads to inexact 
conservation that depends on grid spacing (short dashed lines).}
\label{fig:conserve}
\end{figure*}

Defining
$\Psi\equiv Gh'$, with the prime denoting differentiation with respect to $x$, Taylor
series can be used to show
\begin{equation}
\frac{\Psi_{j+1/2}-\Psi_{j-1/2}}{w_{j}} = \Psi'_{j}\frac{\Delta x_{j}}{w_{j}} + \mathcal{O}\left(\frac{\left(\Delta x\right)_{j}^{2}}{w_{j}}\right ),
\label{eqn:fderr}
\end{equation}
where $\Delta x_{j}=x_{j+1/2}-x_{j-1/2}$.  With the exception of pitch angles
corresponding to trapped particles,~\cite{barnesPoP08,kotschCPC95} the grid points
$\{x_{j}\}$ and associated integration weights $\{w_{j}\}$ in GS2
are chosen according to Gauss-Legendre quadrature rules.~\cite{hilde}  For this case, we show 
numerically in Fig.~\ref{fig:errfd} that 
\begin{equation}
\frac{1}{N}\sum_{j=1}^{N}\frac{\Delta x_{j}}{w_{j}}
=1+ \mathcal{O}\left(\frac{1}{N}\sum_{j=1}^{N}\Delta x_{j}\right)
=1 + \mathcal{O}\left(\frac{1}{N}\right),
\end{equation}
and
\begin{equation}
\max_{j= 2,...,N-1}\left| 1 - \frac{\Delta x_{j}}{w_{j}} \right| = \mathcal{O}\left(\frac{1}{N}\right),
\end{equation}
where $N$ is the number of grid points in $x$.

The boundary points ($j=1,N$) are excluded from the $\max$ operator above.  This
is because $\Delta x / w$ (the factor multiplying $\Psi_{j}'$ in Eqn.~(\ref{eqn:fderr}))
converges to approximately 1.2 for the boundary points 
as the grid spacing is decreased (Fig.~\ref{fig:errfd}).  For the energy diffusion operator, we can make use of
the property that $G(x)=G(x)'=0$ at $x=0$ and $x=\infty$ to show
\begin{equation}
\pm \frac{\Psi_{j\pm1/2}}{w_{j}} = \Psi_{j}' + \mathcal{O}\left(\frac{\left(\Delta x\right)_{j}^{2}}{w_{j}}\right),
\end{equation}
with the plus sign corresponding to $j=1$ and the minus sign to $j=N$.  This is not true
for the Lorentz operator, so we are forced to accept an approximately twenty percent
magnification of the Lorentz operator amplitude at $\xi=\pm1$ and at the trapped-passing
boundaries.  We find that this relatively 
small error at the boundaries has a negligible effect on measureable (velocity space averaged)
quantities in our simulations.

\section{Numerical tests}

We now proceed to demonstrate the validity of our collision operator implementation.  In
particular, we demonstrate conservation properties, satisfaction of Boltzmann's
$H$-Theorem, efficient smoothing in velocity space, and recovery of theoretically expected
results in both collisional (fluid) and collisionless limits.  While we do not claim that the
suite of tests we have performed is exhaustive, it constitutes a convenient set of numerical
benchmarks that can be used for validating collision operators in gyrokinetics.

\subsection{Homogeneous plasma slab}

We first consider the long wavelength limit of a homogeneous plasma slab with Boltzmann
electrons and no variation along the magnetic field ($k_{\parl}=0$).
The gyrokinetic equation for this system simplifies to
\begin{equation}
\pd{\left(\delta f\right)}{t} \approx C_{GK}^{0}[h],
\end{equation}
which means local density, momentum, and energy should be conserved.  In Fig.~\ref{fig:conserve}
we show numerical results for the time evolution of the local density, momentum, and energy for this system.

Without inclusion of the conserving terms (\ref{eqn:scriptuL})-(\ref{eqn:scripte}), we 
see that density is conserved, as guaranteed by the conservative differencing scheme,
while the momentum and energy decay away over several collision times.
Inclusion of the conserving terms provides us with exact (up to numerical precision)
conservation of number, momentum, and energy.  To illustrate the utility of our
conservative implementation, we also present results from a numerical scheme that does not
make use of Eqns.~(\ref{eqn:vpa})-(\ref{eqn:nue}) and that employs a finite difference
scheme that does not possess discrete versions of the Fundamental Theorem of Calculus
and integration by parts.  Specifically, we consider a first order accurate finite difference
scheme similar to that given by Eqn.~(\ref{eqn:fd}), with the only difference being
that the weights in the denominator are replaced with the local grid spacings.  In this case,
we see that density, momentum, and energy are not exactly conserved (how well they
are conserved depends on velocity space resolution, which is 16 pitch angles and 16 energies
for the run considered here).

\begin{figure} 
\begin{center}
\includegraphics[height=3.3in]{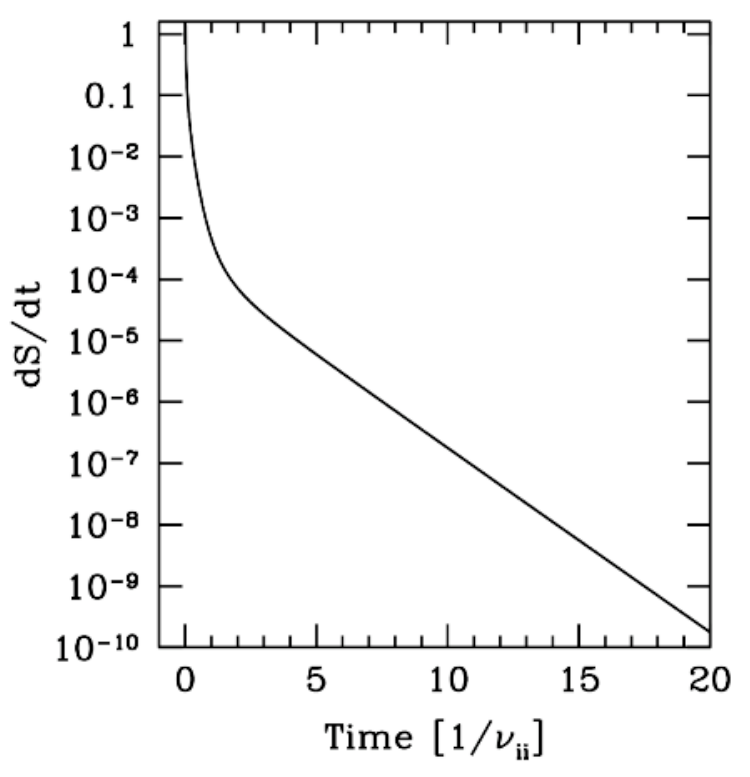}
\end{center}
\caption{Plot of the evolution of entropy generation for the homogeneous plasma slab
over twenty collision times.  Our initial distribution in velocity space is random noise, and 
we use a grid with 16 pitch angles and 8 energies.  The 
entropy generation rate is always nonnegative and approaches zero in the long-time limit.}
\label{fig:htheorem}
\end{figure}

The rate at which our collision operator generates entropy in the homogenous plasma slab 
is shown in Fig.~\ref{fig:htheorem}.  As required by the $H$-Theorem, the rate of entropy 
production is always nonnegative and approaches
zero in the long-time limit as the distribution function approaches a shifted Maxwellian.
We find this to hold independent of both the grid spacing in velocity space and the
initial condition for the distribution function (in Fig.~\ref{fig:htheorem}, the values
of $h(\xi,v)$ were drawn randomly from the uniform distribution on the interval $[-1/2,1/2]$).

\subsection{Resistive damping}

We now modify the system above by adding a finite $A_{\parl}$.  From fluid theory, 
we know that collisional friction between electrons and ions provides
resistivity which leads to the decay of current profiles.  Because the resistive time is long
compared to the collision time, one can neglect $\partial (\delta f) / \partial t$.  However, since
$A_{\parl}\sim k^{-2}$, and we are considering $k\ll 1$, $\partial A_{\parl} / \partial t$ must 
be retained.  The resulting electron equation is of the form of the classical Spitzer problem (see, e.g., Ref.~\onlinecite{helander}):
\begin{equation}
C_{GK}^{0}[h_{e}] = -\frac{eF_{0,e}}{T_{0,e}}\frac{v_{\parl}}{c}\pd{A_{\parl}}{t}.
\end{equation}
The parallel current evolution for this system is given by
\begin{equation}
J_{\parl}(t) = J_{\parl}(t=0)e^{-\eta t},
\label{eqn:djdt}
\end{equation}
where $\eta=1/\sigma_{\parl}$ is the resistivity, $\sigma_{\parl}=1.98 \tau_{e}n_{e}e^{2}/m_{e}c^{2}$ is
the Spitzer conductivity, and $\tau_{e}=3\sqrt{\pi} / 4 \nu_{ei}$ is the electron collision time.

\begin{figure} 
\begin{center}
\includegraphics[height=3.3in]{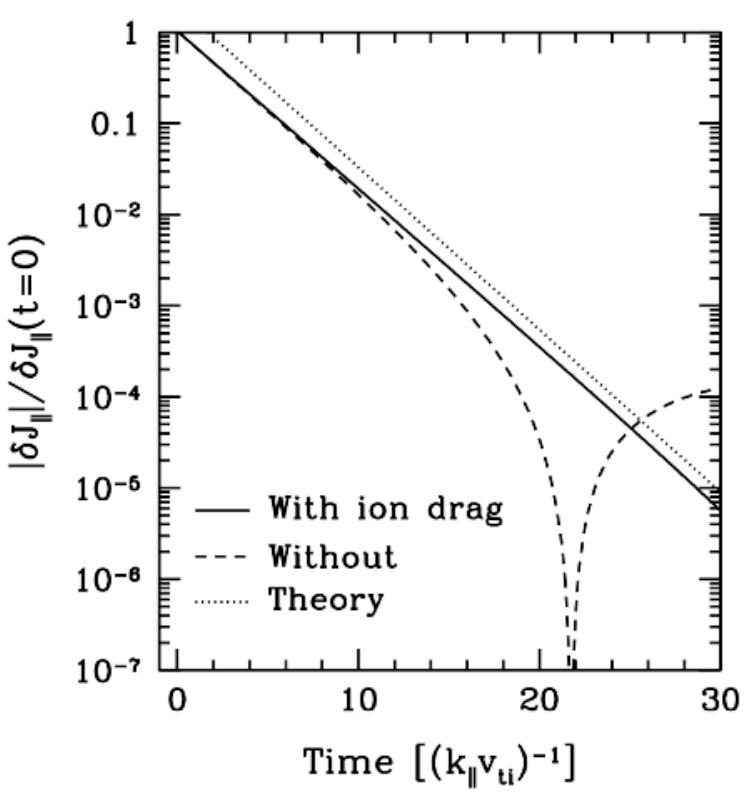}
\end{center}
\caption{Evolution of $|J_{\parl}|$ for the electromagnetic
plasma slab with $\beta=10^{-4}$, $k_{y}\rho_{i}=0.1$, and $\nu_{ei}=10 k_{\parl}v_{th,i}$.  Inclusion
of the ion drag term in the electron-ion collision operator leads to the theoretically predicted 
damping rate for the parallel current given in Eqn.~(\ref{eqn:djdt}) .  Without the ion 
drag term, the parallel current decays past zero (at $t\approx22$) and converges to 
a negative value as the electron flow damps to zero.}
\label{fig:apardamp}
\end{figure}

\begin{figure} 
\begin{center}
\includegraphics[height=3.3in]{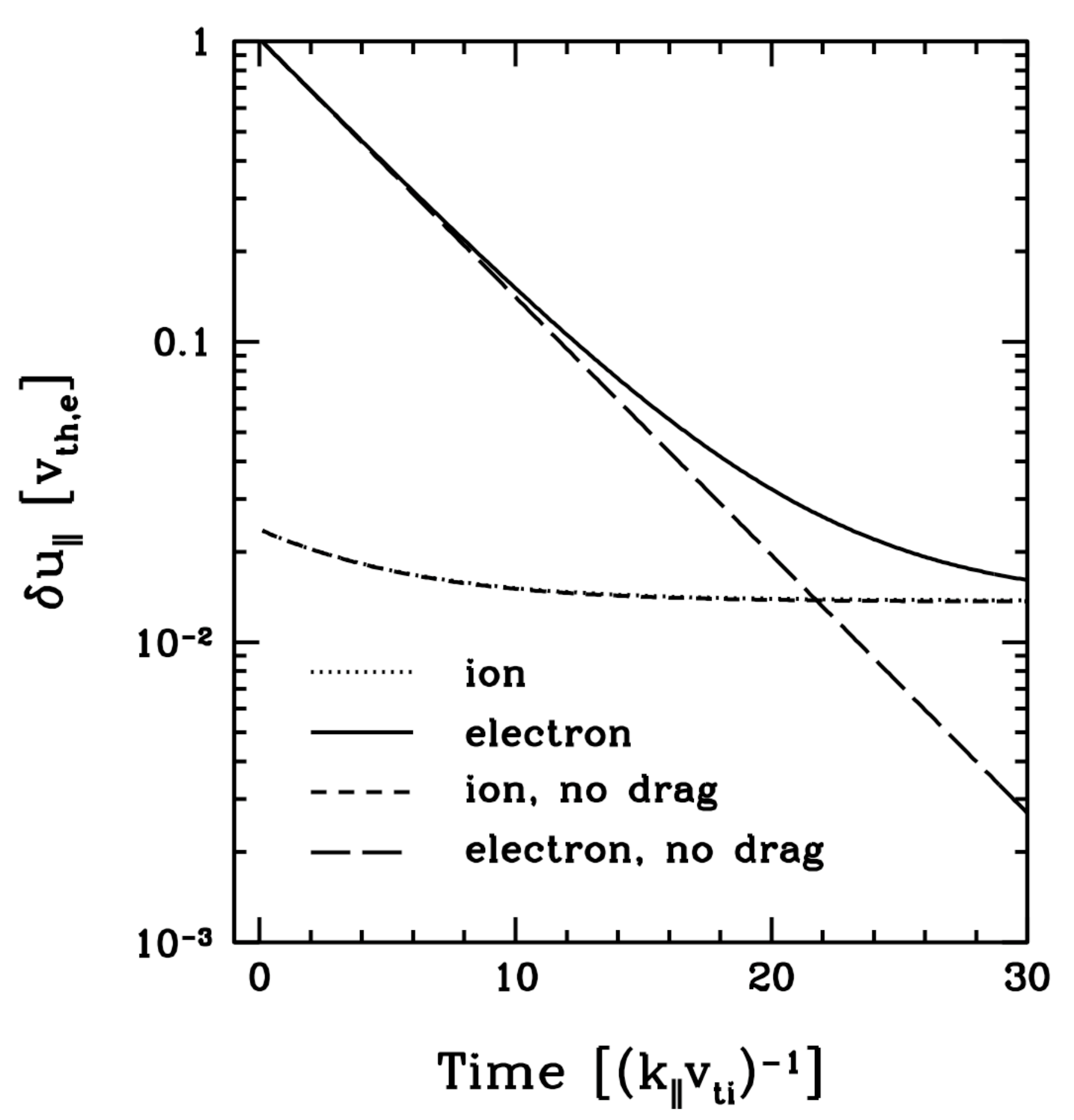}
\end{center}
\caption{Evolution of perturbed parallel flow for the electromagnetic
plasma slab with $\beta=10^{-4}$, $k_{y}\rho_{i}=0.1$, and $\nu_{ei}=10 k_{\parl}v_{ti}$.  
Without inclusion of the ion drag term in Eqn.~(\ref{eqn:Cei}), the electron flow is 
erroneously damped to zero (instead of to the ion flow).}
\label{fig:upardamp}
\end{figure}

We demonstrate that
the numerical implementation of our operator correctly captures this resistive damping
in Figs.~\ref{fig:apardamp} and~\ref{fig:upardamp}.  We also see in these figures that in the 
absence of the ion drag term from Eqn.~(\ref{eqn:Cei}), the electron flow is incorrectly damped 
to zero (instead of to the ion flow), leading to a steady-state current.  

\subsection{Slow mode damping}

We next consider the damping of the slow mode in a homogenous plasma slab
as a function of collisionality.  In the low $k_{\perp}\rho_{i}$, high $\beta_{i}$
limit, one can obtain analytic expressions for the damping rate in both the collisional 
($k_{\parl}\lambda_{mfp}\ll 1$) and collisionless ($k_{\parl}\lambda_{mfp}\gg 1$) 
regimes, where $\lambda_{mfp}$ is the ion mean free path (see e.g. 
Ref.~\onlinecite{schekApJ08}).  The expressions are
\begin{equation}
\omega = \pm k_{\parl}v_{A}\sqrt{1-\left(\frac{\nu_{\parl,i}k_{\parl}}{2v_{A}}\right)^{2}}-i\frac{\nu_{\parl,i}k_{\parl}^{2}}{2}
\label{eqn:collisional}
\end{equation}
for $k_{\parl}\lambda_{mfp}\ll 1$, and
\begin{equation}
\omega = -i\frac{\left|k_{\parl}\right|v_{A}}{\sqrt{\pi\beta_{i}}}
\label{eqn:collisionless}
\end{equation}
for $k_{\parl}\lambda_{mfp}\gg 1$.  Here, $v_{A}=v_{th,i}/\sqrt{\beta_{i}}$ is
the Alfven speed, and $\nu_{\parl,i}$ is the parallel ion viscosity, which is inversely 
proportial to the ion-ion collision frequency, 
$\nu_{ii}$: $\nu_{\parl,i}\propto v_{th,i}^{2}/\nu_{ii}$.  As one would expect, the 
damping in the strongly collisional regime [Eqn.~(\ref{eqn:collisional})] is due primarily to viscosity, while the collisionless
regime [Eqn.~(\ref{eqn:collisionless})] is dominated by Barnes damping.~\cite{barnesPoF66}

\begin{figure} 
\begin{center}
\includegraphics[height=3.3in]{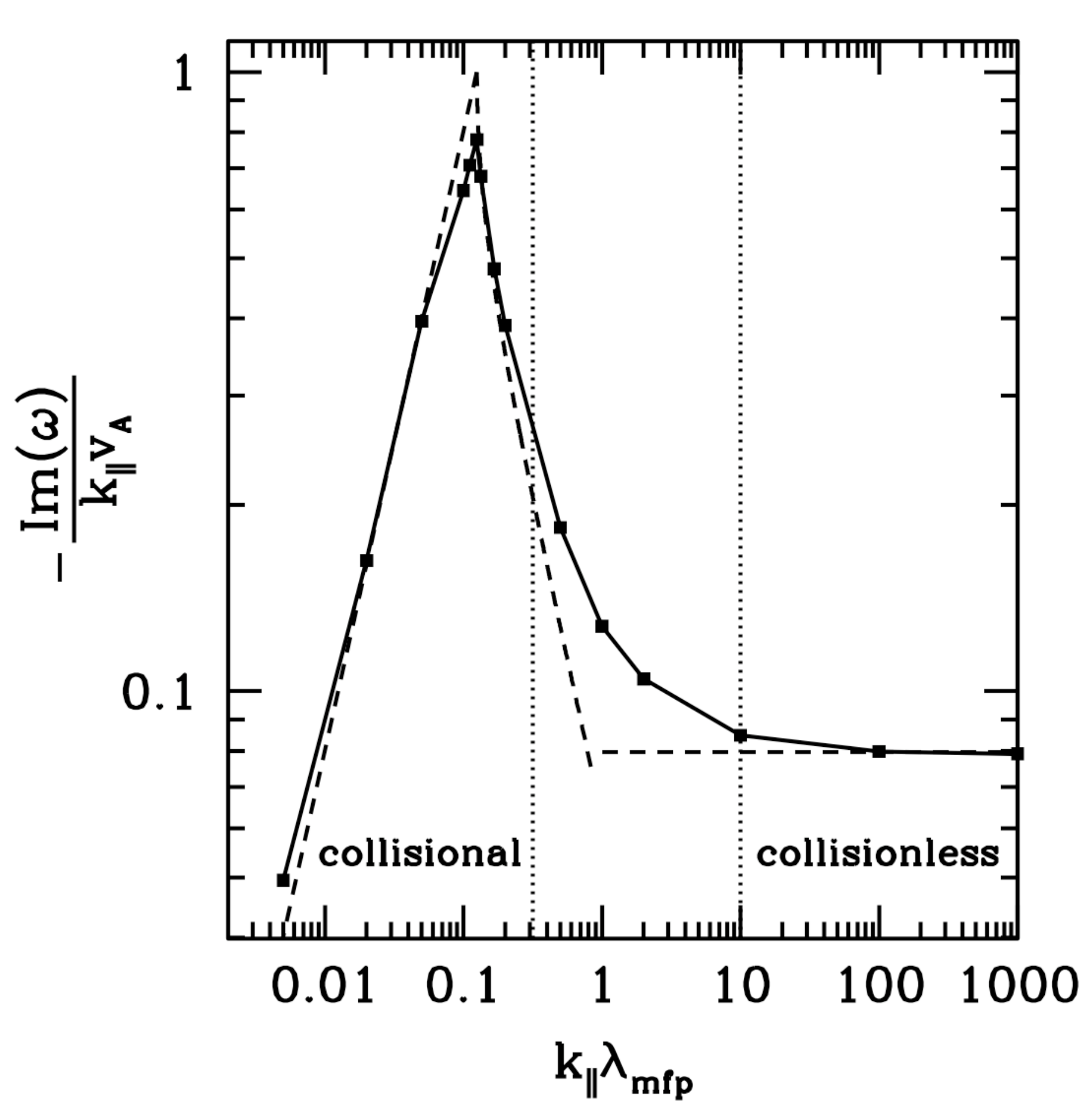}
\end{center}
\caption{Damping rate of the slow mode for a range of collisionalities spanning
the collisionless to strongly collisional regimes.  Dashed lines correspond to the 
theoretical prediction for the damping rate in the collisional ($k_{\parl}\lambda_{mfp}\ll 1$)
and collisionless ($k_{\parl}\lambda_{mfp}\gg 1$) limits.  The solid line is the result 
obtained numerically with GS2.  Vertical dotted lines denote approximate regions 
(collisional and collisionless) for which the analytic theory is valid.}
\label{fig:nuscan}
\end{figure}

In Fig.~\ref{fig:nuscan}, we plot the collisional dependence of the damping rate of the 
slow mode obtained numerically
using the new collision operator implementation in GS2.  In order to isolate the slow mode
in these simulations, we took $\varphi=A_{\parl}=\delta n_{e}=0$ and measured
the damping rate of $\delta B_{\parl}$.  This is possible because $\delta B_{\parl}$
effectively decouples from $\varphi$ and $A_{\parl}$ for our system, and $\delta n_{e}$
can be neglected because $\beta_{i}\gg1$.~\cite{schekApJ08}  
We find quantitative agreement with
the analytic expressions (\ref{eqn:collisional}) and (\ref{eqn:collisionless}) in
the appropriate regimes.  In particular, we recover the correct
viscous behavior in the $k_{\parl}\lambda_{mfp}\ll 1$ limit (damping rate
proportional to $\nu_{\parl,i}$), the correct collisional damping in the $k_{\parl}\lambda_{mfp}\sim 1$
limit (damping rate inversely proportial to $\nu_{\parl,i}$), and the correct collisionless 
(i.e. Barnes) damping in the $k_{\parl}\lambda_{mfp}\gg 1$ limit.

\subsection{Electrostatic turbulence}

Finally, we illustrate the utility of our collision operator in a nonlinear simulation of
electrostatic turbulence in a Z-pinch field configuration.~\cite{freidbergMHD}  We consider the Z-pinch
because it contains much of the physics of toroidal configurations (i.e. curvature)
without some of the complexity (no particle trapping).  At relatively weak pressure gradients,
the dominant gyrokinetic linear instability in the Z-pinch is the entropy 
mode,~\cite{kadomtsevSP60,kesnerPoP00,simakovPoP01,simakovPoP02,kesnerPoP02,ricciPoP06}
which is nonlinearly unstable to secondary instabilities such as Kelvin-Helmholtz.~\cite{ricciPRL06}

In previous numerical investigations of linear~\cite{ricciPoP06} and nonlinear~\cite{ricciPRL06}
plasma dynamics in a Z-pinch, collisions were found to play an important role in the damping
of zonal flows and in providing an effective energy cutoff at short wavelengths.  
However, as pointed out in Ref.~\onlinecite{ricciPRL06}, the Lorentz collision operator 
used in those investigations provided insufficient damping of short wavelength structures
to obtain steady-state fluxes.  Consequently, a model hyper-viscosity had to be employed.

\begin{figure} 
\begin{center}
\includegraphics[height=3.3in]{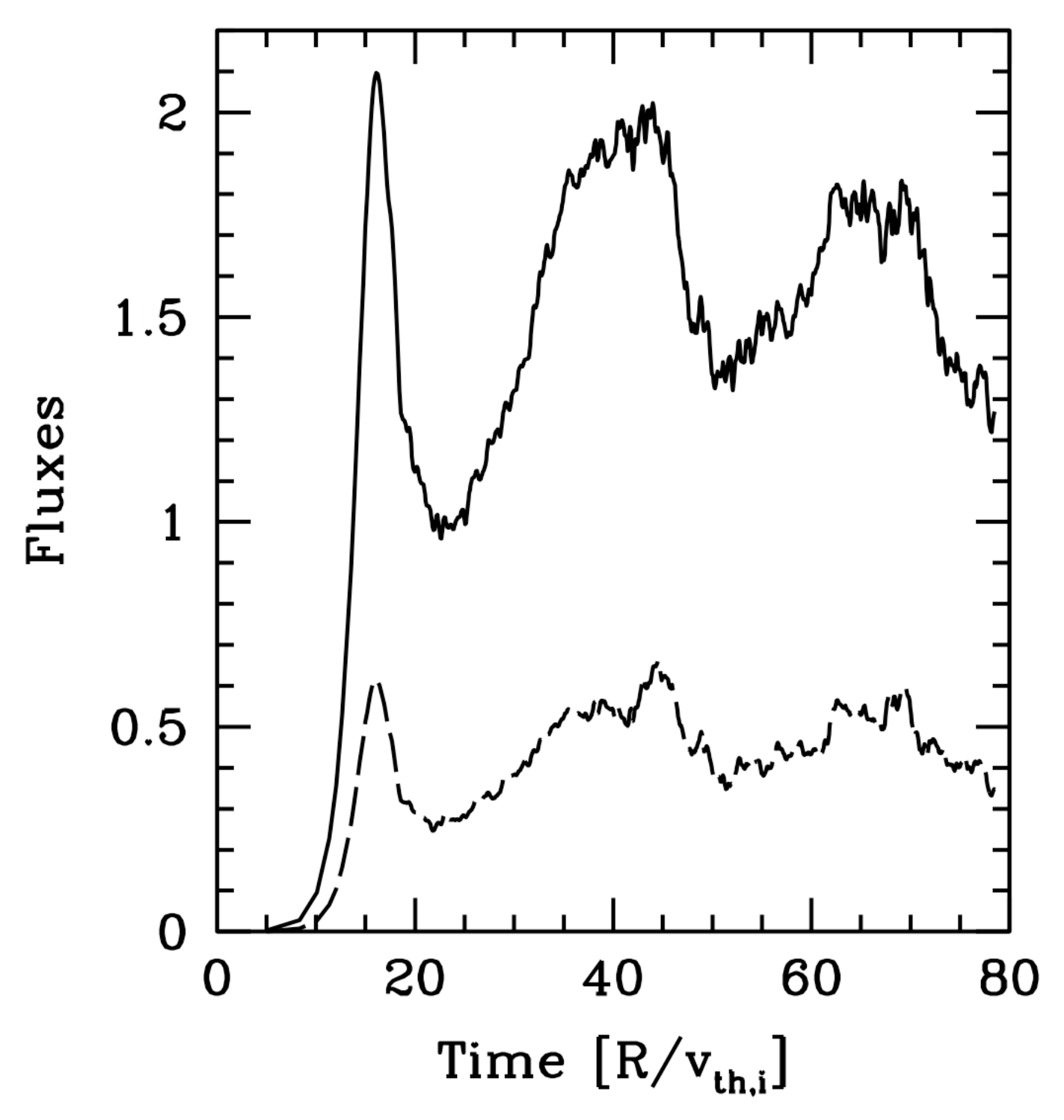}
\end{center}
\caption{Evolution of ion particle and heat fluxes for an electrostatic, 2-species Z-pinch 
simulation.  We are considering $R/L_{n}=2.0$ and $\nu_{ii}=0.01 v_{th,i}/R$.  The particle flux
is indicated by the solid line and is given in units of $(\rho / R) n_{0,i}v_{th,i}$.  The heat 
flux is indicated by the dashed line and is given in units of $(\rho/R) n_{0,i}v_{th,i}^{2}$.
We see that a steady-state is achieved for both fluxes without artificial dissipation.}
\label{fig:nlzp}
\end{figure}

\begin{figure} 
\begin{center}
\includegraphics[height=3.3in]{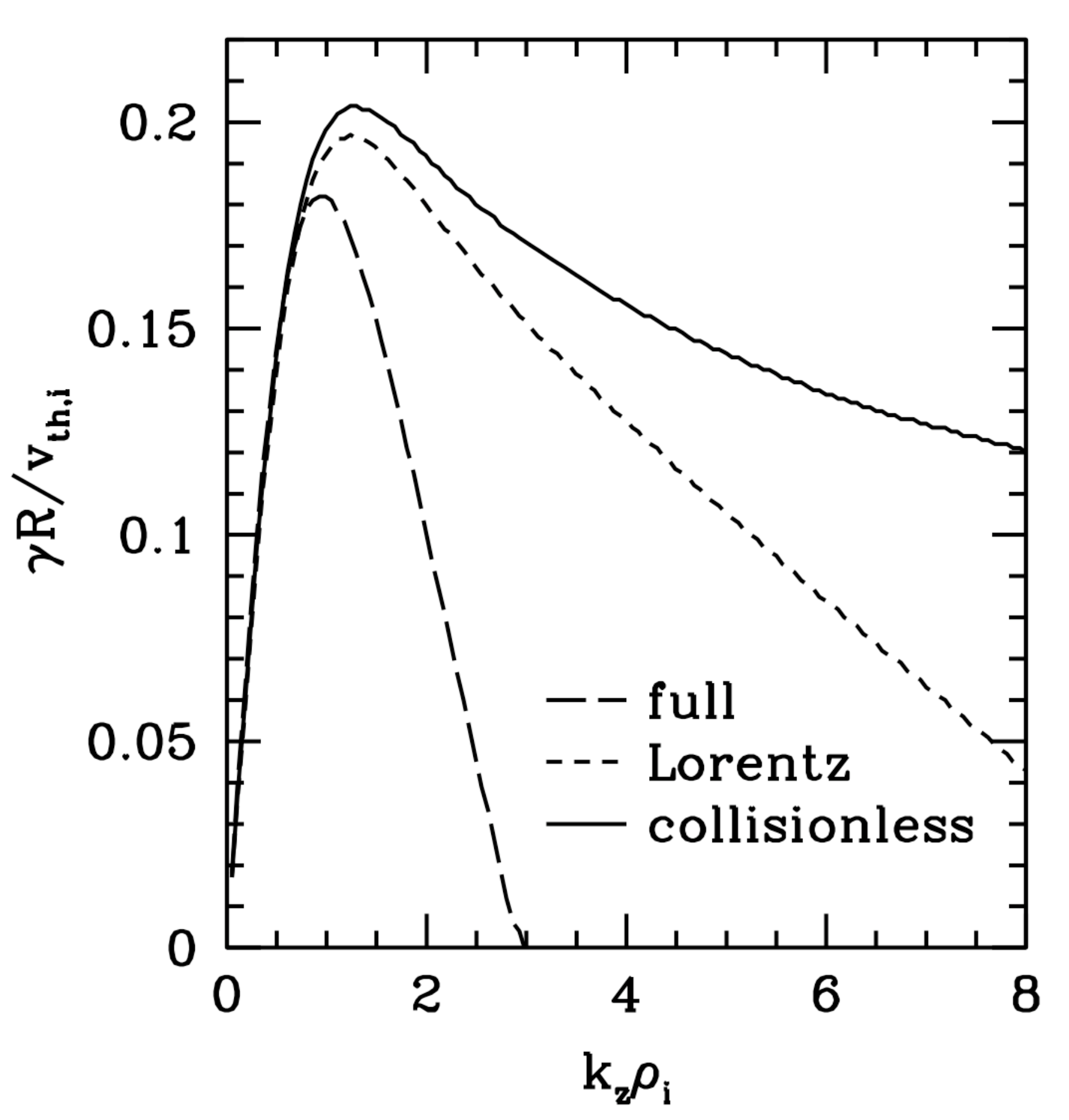}
\end{center}
\caption{Linear growth rate spectrum of the entropy mode in a Z-pinch for $R/L_{n}=2.0$, 
where $R$ is major radius and $L_{n}$ is density gradient scale length.  The solid line
is the collisionless result, and the two dashed lines represent the result of including collisions.
The short dashed line corresponds to using only the Lorentz operator, while the long dashed
line corresponds to using our full model collision operator.  Both collisional cases were
carried out with $\nu_{ii} = 0.01 v_{th,i}/R$.}
\label{fig:lzp}
\end{figure}

We have reproduced a simulation from Ref.~\onlinecite{ricciPRL06} using our new
collison operator, and we find that hyper-viscosity is no longer necessary to obtain
steady-state fluxes (Fig.~\ref{fig:nlzp}).  This can be understood by examining the linear growth
rate spectrum of Fig.~\ref{fig:lzp}.  We see that in this system energy diffusion is much
more efficient at suppressing short wavelength structures than pitch-angle scattering.
Consequently, no artificial dissipation of short wavelength structures is necessary.

\section{Summary}

In Sec.~I we proposed a set of key properties that an ideal dissipation scheme
for gyrokinetics should satisfy.  Namely, the scheme should: limit the scale size of
structures in phase space in order to guarantee the validity of the gyrokinetic ordering 
and to provide numerical resolution at reasonable expense;
conserve particle number, momentum, and energy; and satisfy Boltzmann's $H$-Theorem.
While commonly employed simplified collision operators or
hyperviscosity operators may be adequate for some calculations,~\cite{dimitsPoP00} 
it is important to be able to use the more complete collision operator described in this
paper, which preserves all of these desirable dissipation properties.

In Sec.~II we presented the model collision operator derived in Paper I and discussed
some of its features that strongly influence our choice of numerical implementation.
In particular, we noted that local conservation properties are guaranteed as long as the
($1$, $v_{\parl}$, $v^{2}$) moments of the $k\rho=0$
component of the gyroaveraged collision operator vanish.  Further, we argued that
the collision operator should be treated implicitly because in some regions of phase space,
its amplitude can be large even at very small collisionalities.

Our numerical implementation of the collision operator was described in Sec. III.
We separate collisional and collisionless physics through the use of Godunov 
dimensional splitting and advance the collision operator in time using a backwards
Euler scheme.  The test particle part of the collision operator is differenced using
a scheme that possesses discrete versions of the Fundamental Theorem of Calculus
and integration by parts (upon double application).  These properties are necessary in order 
to exactly satisfy the desired conservation properties in the long wavelength limit.  The
field particle response is treated implicitly with little additional computational expense
by employing repeated application of the Sherman-Morrison formula, as detailed
in Appendix A.

In Sec. IV we presented numerical tests to demonstrate that our implemented
collision operator possesses the properties required for a good gyrokinetic 
dissipation scheme.  In addition to these basic properties, we showed that the
implemented collision operator allows us to correctly capture physics phenomena ranging
from the collisionless to the strongly collisional regimes.  In particular, we provided
examples for which we are able to obtain quantitatively correct results for 
collisionless (Landau or Barnes), resistive, and viscous damping.  

In conclusion, we note that resolution of the collisionless (and collisional) physics in our
simulations was 
obtained solely with physical collisions; no recourse to any form of artificial numerical 
dissipation was necessary.

\begin{acknowledgments} 
We thank S. C. Cowley, R. Numata, and I. Broemstrup for useful discussions.  
M.B., W.D., and P.R. were supported by the US DOE Center for Multiscale Plasma Dynamics.  
I.G.A.\ was supported by a CASE EPSRC studentship in association with UKAEA Fusion 
(Culham).  A.A.S. was supported by an STFC (UK) Advanced Fellowship and STFC Grant 
ST/F002505/1. M.B., G.W.H., and W.D.\ would also like to thank the Leverhulme Trust (UK) 
International Network for Magnetized Plasma Turbulence for travel support.
\end{acknowledgments}

\section{Appendix A: Sherman-Morrison Formulation}

The repeated application of the Sherman-Morrison formula considered here
is an extension of the scheme presented in Tatsuno and Dorland.~\cite{tatsunoAN08} 
Throughout this calculation, we adopt general notation applicable to both
Eqns.~(\ref{eqn:Lsplit}) and (\ref{eqn:DMEsplit}) and provide specific variable definitions in 
Table~\ref{tab:sm}.  
Both Eqns.~(\ref{eqn:Lsplit}) and (\ref{eqn:DMEsplit}) can be written in the form
\begin{equation}
A\mbf{x}=\mbf{b}.
\end{equation}  
\begin{table}
\begin{center}
\begin{tabular} {| c || c | c |}
\hline
variable & $L$ & $D$ \\
\hline \hline
$A$ & $\ 1-\Delta t\left(L+U_L\right) \ $ & $\ 1-\Delta t\left(D+U_D+E\right) \ $\\
\hline
$\mbf{x}$ & $h^{**}$ & $h^{n+1}$ \\
\hline
$\mbf{b}$ & $h^{*}$ & $h^{**}$\\
\hline
$A_{0}$ & $1-\Delta t L$ & $1-\Delta t D$\\
\hline
$\mbf{v}_{0}$ & $\nu_{D}v_{\perp}J_{1}$ & $-\Delta \nu v_{\perp}J_{1}$ \\
\hline
$\mbf{v}_{1}$ & $\nu_{D}v_{\parl}J_{0}$ & $-\Delta \nu v_{\parl}J_{0}$ \\
\hline
$\mbf{v}_{2}$ & $v_{\parl}$ (electrons) & $\nu_{E}v^{2}J_{0}$ \\
& $0$ (ions) &\\
\hline
$\mbf{u}_{0}$ & $-\Delta t\mbf{v}_0/d_{u}$ & $-\Delta t\mbf{v}_0/d_{u}$ \\
\hline
$\mbf{u}_{1}$ & $-\Delta t\mbf{v}_1/d_{u}$ & $-\Delta t\mbf{v}_1/d_{u}$ \\
\hline
$\mbf{u}_{2}$ & $-\Delta t\nu_{D}^{ei}v_{\parl}/d_{q}$ (electrons) & $-\Delta t\mbf{v}_2/d_{q}$ \\
& $0$ (ions) &\\
\hline
$d_{u}$ & $\int d^{3}v \ \nu_{D} v_{\parl}^{2} F_{0}$ & $\int d^{3}v \ \Delta\nu v_{\parl}^{2} F_{0}$ \\
\hline
$d_{q}$ & $v_{th,e}^{2}/2$ & $\int d^{3}v \ \nu_{E} v^{4} F_{0}$ \\
\hline
\end{tabular}
\end{center}
\caption{Sherman-Morrison variable definitions for Lorentz and energy diffusion
operator equations}
\label{tab:sm}
\end{table}
Because $\lb \mathcal{U} \rb$ and $\lb \mathcal{E} \rb$ are integral operators, we can write
them as tensor products so that
\begin{equation}
A\equiv A_{0}+\mbf{u}_{0}\otimes\mbf{v}_{0}+\mbf{u}_{1}\otimes\mbf{v}_{1}
+\mbf{u}_{2}\otimes\mbf{v}_{2}.
\end{equation}
We now define 
\begin{equation}
A_{1} = A_{0}+\mbf{u}_{0}\otimes\mbf{v}_{0}, \ \ A_{2}=A_{1}+\mbf{u}_{1}\otimes\mbf{v}_{1},
\end{equation}
so that
\begin{equation}
\left(A_{2}+\mbf{u}_{2}\otimes\mbf{v}_{2}\right)\mbf{x} = \mbf{b}.
\end{equation}
Applying the Sherman-Morrison formula to this equation, we find
\begin{equation}
\mbf{x} = \mbf{y}_{2} - \frac{\mbf{v}_{2}\cdot\mbf{y}_{2}}{1+\mbf{v}_{2}\cdot\mbf{z}_{2}}\mbf{z}_{2},
\end{equation}
where $A_{2}\mbf{y}_{2}=\mbf{b}$ and $A_{2}\mbf{z}_{2}=\mbf{u}_{2}$.

Applying the Sherman-Morrison formula to each of these equations gives
\begin{gather}
\mbf{y}_{2} = \mbf{y}_{1} - \frac{\mbf{v}_{1}\cdot\mbf{y}_{1}}{1+\mbf{v}_{1}\cdot\mbf{w}_{1}}\mbf{w}_{1}\\
\mbf{z}_{2} = \mbf{z}_{1} - \frac{\mbf{v}_{1}\cdot\mbf{z}_{1}}{1+\mbf{v}_{1}\cdot\mbf{w}_{1}}\mbf{w}_{1},
\end{gather}
where $A_{1}\mbf{y}_{1}=\mbf{b}$, $A_{1}\mbf{w}_{1}=\mbf{u}_{1}$, and 
$A_{1}\mbf{z}_{1}=\mbf{u}_{2}$.  A final application of Sherman-Morrison to 
these three equations yields
\begin{gather}
\mbf{y}_{1} = \mbf{y}_{0} - \frac{\mbf{v}_{0}\cdot\mbf{y}_{0}}{1+\mbf{v}_{0}\cdot\mbf{s}_{0}}\mbf{s}_{0}\\
\mbf{w}_{1} = \mbf{w}_{0} - \frac{\mbf{v}_{0}\cdot\mbf{w}_{0}}{1+\mbf{v}_{0}\cdot\mbf{s}_{0}}\mbf{s}_{0}\\
\mbf{z}_{1} = \mbf{z}_{0} - \frac{\mbf{v}_{0}\cdot\mbf{z}_{0}}{1+\mbf{v}_{0}\cdot\mbf{s}_{0}}\mbf{s}_{0},
\end{gather}
where $A_{0}\mbf{y}_{0}=\mbf{b}$, $A_{0}\mbf{s}_{0}=\mbf{u}_{0}$,
$A_{0}\mbf{w}_{0}=\mbf{u}_{1}$, and $A_{0}\mbf{z}_{0}=\mbf{u}_{2}$.

We can simplify our expressions by noting that $\mbf{v}_{0,1,2}$ and $\mbf{u}_{0,1,2}$ have
definite parity in $v_{\parl}$.  A number of inner products then vanish by symmetry, leaving
the general expressions
\begin{gather}
\mbf{y}_{2} = \mbf{y}_{0} - \left[\frac{\mbf{v}_{0}\cdot\mbf{y}_{0}}{1+\mbf{v}_{0}\cdot\mbf{s}_{0}}\right]\mbf{s}_{0} - \left[\frac{\mbf{v}_{1}\cdot\mbf{y}_{0}}{1+\mbf{v}_{1}\cdot\mbf{w}_{0}}\right]\mbf{w}_{0}\\
\mbf{z}_{2} = \mbf{z}_{0} - \left[\frac{\mbf{v}_{0}\cdot\mbf{z}_{0}}{1+\mbf{v}_{0}\cdot\mbf{s}_{0}}\right]\mbf{s}_{0}- \left[\frac{\mbf{v}_{1}\cdot\mbf{z}_{0}}{1+\mbf{v}_{1}\cdot\mbf{w}_{0}}\right]\mbf{w}_{0}.
\end{gather}

\section{Appendix B: Compact differencing of the test-particle operator}

In this appendix, we derive a second order accurate compact difference scheme for pitch-angle 
scattering and energy diffusion on an unequally spaced grid.  The higher order of accuracy
of this scheme is desirable, but it does not possess discrete versions of the Fundamental
Theorem of Calculus and integration by parts when used with Gauss-Legendre quadrature
(or any other integration scheme with grid spacings unequal to integration weights).
Consequently, one should utilize this scheme only if integration weights and grid spacings 
are equal, or if exact satisfaction of conservation properties is not considered important.

For convenience, we begin by noting that Eqns.~(\ref{eqn:Lsplit}) and 
(\ref{eqn:DMEsplit}) can both be written in the general form
\begin{equation}
\left(\pd{h}{t}\right)_{C} = H\left(Gh'\right)' + S=H\left(G'h'+Gh''\right)+S,
\label{eqn:cmpctgk}
\end{equation}
where: for the Lorentz operator equation (\ref{eqn:Lsplit}) we identify $H=1$, 
$G=\nu_{D}\left(1-\xi^{2}\right)/2$,
$S=U_{L}[h]-k^{2}v^{2}\nu_{D}\left(1+\xi^{2}\right)h/4\Omega_{0}^{2}$,
and the prime denotes differentiation with respect to $\xi$; and for the energy diffusion
operator equation (\ref{eqn:DMEsplit}), we identify $H=1/2v^{2}F_{0}$,
$G=\nu_{\parl}v^{4}F_{0}$, $S=U_{D}[h]+E[h]-k^{2}v^{2}\nu_{\parl}\left(1-\xi^{2}\right)h/4\Omega_{0}^{2}$,
and the prime denotes differentiation with respect to $v$.  Here, the $h$ we are using
is actually normalized by $F_{0}$.

Employing Taylor Series expansions of $h$, we obtain the expressions
\begin{equation}
h_{i}' = \frac{\delta_{-}^{2}\left(h_{i+1}-h_{i}\right) + \delta_{+}^{2}\left(h_{i}-h_{i-1}\right)}{\delta_{+}\delta_{-}\left(\delta_{+}+\delta_{-}\right)} + \mathcal{O}[\delta^{2}],
\end{equation}
and
\begin{equation}
h_{i}'' = 2\frac{\delta_{-}\left(h_{i+1}-h_{i}\right) - \delta_{+}\left(h_{i}-h_{i-1}\right)}{\delta_{+}\delta_{-}\left(\delta_{+}+\delta_{-}\right)}+\frac{\delta_{-}-\delta_{+}}{3}h_{i}''' + \mathcal{O}[\delta^{2}],
\label{eqn:hpp}
\end{equation}
where $i$ denotes evaluation at the velocity space gridpoint $x_{i}$, and $\delta_{\pm}\equiv \left|x_{i\pm 1}-x_{i}\right|$ (here $x$ is a dummy
variable representing either $\xi$ or $v$).  In order for the $h_{i}''$ expression to be second order accurate, we must obtain
a first order accurate expression for $h_{i}'''$ in terms of $h_{i}$, $h_{i}'$, and $h_{i}''$.  We accomplish this by differentiating Eqn.~(\ref{eqn:cmpctgk}) with 
respect to $x$:
\begin{equation}
\left(\pd{h'}{t}\right)_{C} = H'\left(Gh'\right)' + H\left(Gh'\right)''+ S'
\label{eqn:hp}
\end{equation}
and rearranging terms to find
\begin{equation}
\begin{split}
h_{i}''' &= \frac{1}{H_{i}G_{i}}\Big[\left(\pd{h_{i}}{t}\right)_{C} - H_{i}'\left(G_{i}'h_{i}' + G_{i}h_{i}''\right)\\
&-H_{i}\left(G_{i}''h_{i}+2G_{i}'h_{i}''\right)-S_{i}'\Big] + \mathcal{O}[\delta],
\end{split}
\end{equation}
where, unless denoted otherwise, all quantities are taken at the $n+1$ time level.
Plugging this result into Eqn.~\ref{eqn:hpp} and grouping terms, we have
\begin{equation}
\begin{split}
\mu_{i}h_{i}''& =  \frac{\delta_{-}-\delta_{+}}{3H_{i}G_{i}} \left[h_{i}'\left(\frac{1}{\Delta t}-H_{i}'G_{i}'-H_{i}G_{i}''\right) - \frac{\left(h_{i}^{n}\right)'}{\Delta t}-S_{i}'\right] \\
&+ 2\frac{\delta_{-}\left(h_{i+1}-h_{i}\right)-\delta_{+}\left(h_{i}-h_{i-1}\right)}{\delta_{+}\delta_{-}\left(\delta_{+}+\delta_{-}\right)}+ \mathcal{O}[\delta^{2}],
\end{split}
\end{equation}
where $\mu_{i}=1+\left(\delta_{+}-\delta_{-}\right)\left(H_{i}'G_{i}+2H_{i}G_{i}'\right)/3H_{i}G_{i}$,
and we have taken $(\partial h/\partial t)_{C} = (h^{n+1}-h^{n})/\Delta t$.  Using Eqn.~(\ref{eqn:hp})
and the above result in Eqn.~(\ref{eqn:cmpctgk}), we find
\begin{equation}
\begin{split}
\left(\pd{h}{t}\right)_{C} &= h_{i}'\left(H_{i}G_{i}' - \frac{\delta_{+}+\delta_{-}}{3\mu_{i}}\left[\frac{1}{\Delta t} - H_{i}'G_{i}' - H_{i}G_{i}''\right]\right)\\
&+\frac{H_{i}G_{i}}{\mu_{i}}\left(2\frac{\delta_{-}\left(h_{i+1}-h_{i}\right) - \delta_{+}\left(h_{i}-h_{i-1}\right)}{\delta_{+}\delta_{-}\left(\delta_{+}+\delta_{-}\right)}\right.\\
&+\left.\frac{\delta_{+}+\delta_{-}}{3H_{i}G_{i}}\left[\frac{\left(h_{i}^{n}\right)'}{\Delta t} + S_{i}'\right]\right) + S_{i}+\mathcal{O}[\delta^{2}].
\label{eqn:cmpctgen}
\end{split}
\end{equation}This is the general compact differenced form to be used when solving Eqns.~(\ref{eqn:Lsplit}) and (\ref{eqn:DMEsplit}).

In order to illustrate how compact differencing affects the implicit solution using Sherman-Morrison,
we present the result of using the particular form of $S$ for energy diffusion in Eqn.~(\ref{eqn:cmpctgen}):
\begin{equation}
\begin{split}
\frac{h_{i}^{n+1}-h_{i}^{n}}{\Delta t}& = \frac{h_{i+1}}{\delta_{+}\left(\delta_{+}+\delta_{-}\right)}\left(\frac{2H_{i}G_{i}}{\mu_{i}}+\delta_{-}\zeta_{i}\right)\\
&\frac{h_{i-1}}{\delta_{-}\left(\delta_{+}+\delta_{-}\right)}\left(\frac{2H_{i}G_{i}}{\mu_{i}}-\delta_{+}\zeta_{i}\right)\\
&\frac{h_{i}}{\delta_{-}\delta_{+}}\left(-\frac{2H_{i}G_{i}}{\mu_{i}}+\left(\delta_{+}-\delta_{-}\right)\zeta_{i}-\delta_{+}\delta_{-}K\tilde{\nu}_{s}\right)\\
&\frac{\sigma_{i}}{\Delta t}\Bigg(h_{i+1}^{n}\frac{\delta_{-}}{\delta_{+}\left(\delta_{+}+\delta_{-}\right)} - h_{i-1}^{n}\frac{\delta_{+}}{\delta_{-}\left(\delta_{+}+\delta_{-}\right)}\\
&+h_{i}^{n}\frac{\delta_{+}-\delta_{-}}{\delta_{+}\delta_{-}}\Bigg) + \tilde{U}_{\parl}V_{\parl} + \tilde{U}_{\perp}V_{\perp} + \tilde{U}_{q}q + \mathcal{O}[\delta^{2}],
\label{eqn:cmpctEdiff}
\end{split}
\end{equation}
where 
\begin{gather*}
\sigma_{i} = \left(\delta_{+}-\delta_{-}\right)/3\mu_{i}\\
K=k^{2}v_{th}^{2}\left(1-\xi^{2}\right)/8\Omega_{0}^{2}\\
\zeta_{i}=H_{i}G_{i}'-\sigma_{i} \left(1/\Delta t - H_{i}'G_{i}' -H_{i}G_{i}'' + K\nu_{s}\right)\\
\nu_{s}=\nu_{\parl}v_{th}^{2}/2v^{2}\\
\tilde{A}=A+\sigma A'\\
U_{\perp,\parl,q}=\mbf{u}_{0,1,2}\\
V_{\perp,\parl,q}=\mbf{v}_{0,1,2}
\end{gather*}
with $\mbf{u}$ and $\mbf{v}$ given in Table~\ref{tab:sm}.

We see
that the only significant effects of compact differencing on numerical implementation are: 
modification of $h^{**}$ in Eqn.~(\ref{eqn:DMEsplit}) to reflect the $h^{n}$ terms
on the right-hand side of Eqn.~(\ref{eqn:cmpctEdiff}); and modification
of the $U_{\parl}$, $U_{\perp}$, and $U_{q}$ terms that appear in Sherman-Morrison 
($\mbf{u}_{0,1,2}$ from Appendix A) to include an additional $\sigma U'$ term.


\begin{thebibliography}{51}
\expandafter\ifx\csname natexlab\endcsname\relax\def\natexlab#1{#1}\fi
\expandafter\ifx\csname bibnamefont\endcsname\relax
  \def\bibnamefont#1{#1}\fi
\expandafter\ifx\csname bibfnamefont\endcsname\relax
  \def\bibfnamefont#1{#1}\fi
\expandafter\ifx\csname citenamefont\endcsname\relax
  \def\citenamefont#1{#1}\fi
\expandafter\ifx\csname url\endcsname\relax
  \def\url#1{\texttt{#1}}\fi
\expandafter\ifx\csname urlprefix\endcsname\relax\def\urlprefix{URL }\fi
\providecommand{\bibinfo}[2]{#2}
\providecommand{\eprint}[2][]{\url{#2}}

\bibitem[{\citenamefont{Bolton and Ware}(1983)}]{boltonPoF83}
\bibinfo{author}{\bibfnamefont{C.}~\bibnamefont{Bolton}} \bibnamefont{and}
  \bibinfo{author}{\bibfnamefont{A.~A.} \bibnamefont{Ware}},
  \bibinfo{journal}{Phys. Fluids} \textbf{\bibinfo{volume}{26}},
  \bibinfo{pages}{459} (\bibinfo{year}{1983}).

\bibitem[{\citenamefont{Xu}(2008)}]{xuPRE08}
\bibinfo{author}{\bibfnamefont{X.~Q.} \bibnamefont{Xu}},
  \bibinfo{journal}{Phys. Rev. E} \textbf{\bibinfo{volume}{78}},
  \bibinfo{pages}{016406} (\bibinfo{year}{2008}).

\bibitem[{\citenamefont{Ernst et~al.}(2004)\citenamefont{Ernst, Bonoli, Catto,
  Dorland, Fiore, Granetz, Greenwald, Hubbard, Porkolab, Redi
  et~al.}}]{ernstPoP04}
\bibinfo{author}{\bibfnamefont{D.~R.} \bibnamefont{Ernst}},
  \bibinfo{author}{\bibfnamefont{P.~T.} \bibnamefont{Bonoli}},
  \bibinfo{author}{\bibfnamefont{P.~J.} \bibnamefont{Catto}},
  \bibinfo{author}{\bibfnamefont{W.}~\bibnamefont{Dorland}},
  \bibinfo{author}{\bibfnamefont{C.~L.} \bibnamefont{Fiore}},
  \bibinfo{author}{\bibfnamefont{R.~S.} \bibnamefont{Granetz}},
  \bibinfo{author}{\bibfnamefont{M.}~\bibnamefont{Greenwald}},
  \bibinfo{author}{\bibfnamefont{A.~E.} \bibnamefont{Hubbard}},
  \bibinfo{author}{\bibfnamefont{M.}~\bibnamefont{Porkolab}},
  \bibinfo{author}{\bibfnamefont{M.~H.} \bibnamefont{Redi}},
  \bibnamefont{et~al.}, \bibinfo{journal}{Phys. Plasmas}
  \textbf{\bibinfo{volume}{11}}, \bibinfo{pages}{2637} (\bibinfo{year}{2004}).

\bibitem[{\citenamefont{Ernst et~al.}(2006)\citenamefont{Ernst, Basse, Dorland,
  Fiore, Lin, Long, Porkolab, Zeller, and Zhurovich}}]{ernstIAEA06}
\bibinfo{author}{\bibfnamefont{D.~R.} \bibnamefont{Ernst}},
  \bibinfo{author}{\bibfnamefont{N.}~\bibnamefont{Basse}},
  \bibinfo{author}{\bibfnamefont{W.}~\bibnamefont{Dorland}},
  \bibinfo{author}{\bibfnamefont{C.~L.} \bibnamefont{Fiore}},
  \bibinfo{author}{\bibfnamefont{L.}~\bibnamefont{Lin}},
  \bibinfo{author}{\bibfnamefont{A.}~\bibnamefont{Long}},
  \bibinfo{author}{\bibfnamefont{M.}~\bibnamefont{Porkolab}},
  \bibinfo{author}{\bibfnamefont{K.}~\bibnamefont{Zeller}}, \bibnamefont{and}
  \bibinfo{author}{\bibfnamefont{K.}~\bibnamefont{Zhurovich}},
  \bibinfo{journal}{Proc. 21st IAEA Fusion Energy Conf.}
  (\bibinfo{year}{2006}).

\bibitem[{\citenamefont{Kotschenreuther
  et~al.}(1995)\citenamefont{Kotschenreuther, Rewoldt, and Tang}}]{kotschCPC95}
\bibinfo{author}{\bibfnamefont{M.}~\bibnamefont{Kotschenreuther}},
  \bibinfo{author}{\bibfnamefont{G.}~\bibnamefont{Rewoldt}}, \bibnamefont{and}
  \bibinfo{author}{\bibfnamefont{W.~M.} \bibnamefont{Tang}},
  \bibinfo{journal}{Comp. Phys. Comm.} \textbf{\bibinfo{volume}{88}},
  \bibinfo{pages}{128} (\bibinfo{year}{1995}).

\bibitem[{\citenamefont{Federici et~al.}(1987)\citenamefont{Federici, Lee, and
  Tang}}]{federiciPoF87}
\bibinfo{author}{\bibfnamefont{J.~F.} \bibnamefont{Federici}},
  \bibinfo{author}{\bibfnamefont{W.~W.} \bibnamefont{Lee}}, \bibnamefont{and}
  \bibinfo{author}{\bibfnamefont{W.~M.} \bibnamefont{Tang}},
  \bibinfo{journal}{Phys. Fluids} \textbf{\bibinfo{volume}{30}},
  \bibinfo{pages}{425} (\bibinfo{year}{1987}).

\bibitem[{\citenamefont{Rewoldt and Tang}(1990)}]{rewoldtPoF90}
\bibinfo{author}{\bibfnamefont{G.}~\bibnamefont{Rewoldt}} \bibnamefont{and}
  \bibinfo{author}{\bibfnamefont{W.~M.} \bibnamefont{Tang}},
  \bibinfo{journal}{Phys. Fluids B} \textbf{\bibinfo{volume}{2}},
  \bibinfo{pages}{318} (\bibinfo{year}{1990}).

\bibitem[{\citenamefont{Applegate et~al.}(2007)\citenamefont{Applegate, Roach,
  Connor, Cowley, Dorland, Hastie, and Joiner}}]{applegatePPCF07}
\bibinfo{author}{\bibfnamefont{D.~J.} \bibnamefont{Applegate}},
  \bibinfo{author}{\bibfnamefont{C.~M.} \bibnamefont{Roach}},
  \bibinfo{author}{\bibfnamefont{J.~W.} \bibnamefont{Connor}},
  \bibinfo{author}{\bibfnamefont{S.~C.} \bibnamefont{Cowley}},
  \bibinfo{author}{\bibfnamefont{W.}~\bibnamefont{Dorland}},
  \bibinfo{author}{\bibfnamefont{R.~J.} \bibnamefont{Hastie}},
  \bibnamefont{and} \bibinfo{author}{\bibfnamefont{N.}~\bibnamefont{Joiner}},
  \bibinfo{journal}{Plasma Phys. Control. Fusion}
  \textbf{\bibinfo{volume}{49}}, \bibinfo{pages}{1113} (\bibinfo{year}{2007}).

\bibitem[{\citenamefont{Xiao et~al.}(2007)\citenamefont{Xiao, Catto, and
  Dorland}}]{xiaoPoP07}
\bibinfo{author}{\bibfnamefont{Y.}~\bibnamefont{Xiao}},
  \bibinfo{author}{\bibfnamefont{P.~J.} \bibnamefont{Catto}}, \bibnamefont{and}
  \bibinfo{author}{\bibfnamefont{W.}~\bibnamefont{Dorland}},
  \bibinfo{journal}{Phys. Plasmas} \textbf{\bibinfo{volume}{14}},
  \bibinfo{pages}{055910} (\bibinfo{year}{2007}).

\bibitem[{\citenamefont{Krommes and Hu}(1994)}]{krommesPop94}
\bibinfo{author}{\bibfnamefont{J.~A.} \bibnamefont{Krommes}} \bibnamefont{and}
  \bibinfo{author}{\bibfnamefont{G.}~\bibnamefont{Hu}}, \bibinfo{journal}{Phys.
  Plasmas} \textbf{\bibinfo{volume}{1}}, \bibinfo{pages}{3211}
  (\bibinfo{year}{1994}).

\bibitem[{\citenamefont{Krommes}(1999)}]{krommesPoP99}
\bibinfo{author}{\bibfnamefont{J.~A.} \bibnamefont{Krommes}},
  \bibinfo{journal}{Phys. Plasmas} \textbf{\bibinfo{volume}{6}},
  \bibinfo{pages}{1477} (\bibinfo{year}{1999}).

\bibitem[{\citenamefont{Schekochihin et~al.}(2007)\citenamefont{Schekochihin,
  Cowley, Dorland, Hammett, Howes, Quataert, and Tatsuno}}]{schekApJ08}
\bibinfo{author}{\bibfnamefont{A.~A.} \bibnamefont{Schekochihin}},
  \bibinfo{author}{\bibfnamefont{S.~C.} \bibnamefont{Cowley}},
  \bibinfo{author}{\bibfnamefont{W.}~\bibnamefont{Dorland}},
  \bibinfo{author}{\bibfnamefont{G.~W.} \bibnamefont{Hammett}},
  \bibinfo{author}{\bibfnamefont{G.~G.} \bibnamefont{Howes}},
  \bibinfo{author}{\bibfnamefont{E.}~\bibnamefont{Quataert}}, \bibnamefont{and}
  \bibinfo{author}{\bibfnamefont{T.}~\bibnamefont{Tatsuno}},
  \bibinfo{journal}{Astrophys. J. Suppl., submitted}  (\bibinfo{year}{2007}),
  \bibinfo{note}{arXiv: 0704.0044}.

\bibitem[{\citenamefont{Schekochihin et~al.}(2008)\citenamefont{Schekochihin,
  Cowley, Dorland, Hammett, Howes, Plunk, Quataert, and Tatsuno}}]{schekPPCF08}
\bibinfo{author}{\bibfnamefont{A.~A.} \bibnamefont{Schekochihin}},
  \bibinfo{author}{\bibfnamefont{S.~C.} \bibnamefont{Cowley}},
  \bibinfo{author}{\bibfnamefont{W.}~\bibnamefont{Dorland}},
  \bibinfo{author}{\bibfnamefont{G.~W.} \bibnamefont{Hammett}},
  \bibinfo{author}{\bibfnamefont{G.~G.} \bibnamefont{Howes}},
  \bibinfo{author}{\bibfnamefont{G.~G.} \bibnamefont{Plunk}},
  \bibinfo{author}{\bibfnamefont{E.}~\bibnamefont{Quataert}}, \bibnamefont{and}
  \bibinfo{author}{\bibfnamefont{T.}~\bibnamefont{Tatsuno}},
  \bibinfo{journal}{Plasma Phys. Control. Fusion, in press}
  (\bibinfo{year}{2008}), \bibinfo{note}{arXiv: 0806.1069}.

\bibitem[{\citenamefont{Barnes and Dorland}(2008)}]{barnesPoP08}
\bibinfo{author}{\bibfnamefont{M.}~\bibnamefont{Barnes}} \bibnamefont{and}
  \bibinfo{author}{\bibfnamefont{W.}~\bibnamefont{Dorland}},
  \bibinfo{journal}{Phys. Plasmas, submitted}  (\bibinfo{year}{2008}).

\bibitem[{\citenamefont{Grant and Feix}(1967)}]{grantPoF67}
\bibinfo{author}{\bibfnamefont{F.~C.} \bibnamefont{Grant}} \bibnamefont{and}
  \bibinfo{author}{\bibfnamefont{M.~R.} \bibnamefont{Feix}},
  \bibinfo{journal}{Phys. Fluids} \textbf{\bibinfo{volume}{10}},
  \bibinfo{pages}{696} (\bibinfo{year}{1967}).

\bibitem[{\citenamefont{Nevins et~al.}(2005)\citenamefont{Nevins, Hammett,
  Dimits, Dorland, and Shumaker}}]{nevinsPoP05}
\bibinfo{author}{\bibfnamefont{W.}~\bibnamefont{Nevins}},
  \bibinfo{author}{\bibfnamefont{G.~W.} \bibnamefont{Hammett}},
  \bibinfo{author}{\bibfnamefont{A.~M.} \bibnamefont{Dimits}},
  \bibinfo{author}{\bibfnamefont{W.}~\bibnamefont{Dorland}}, \bibnamefont{and}
  \bibinfo{author}{\bibfnamefont{D.~E.} \bibnamefont{Shumaker}},
  \bibinfo{journal}{Phys. Plasmas} \textbf{\bibinfo{volume}{12}},
  \bibinfo{pages}{122305} (\bibinfo{year}{2005}).

\bibitem[{\citenamefont{Frieman and Chen}(1982)}]{friemanPoF82}
\bibinfo{author}{\bibfnamefont{E.~A.} \bibnamefont{Frieman}} \bibnamefont{and}
  \bibinfo{author}{\bibfnamefont{L.}~\bibnamefont{Chen}},
  \bibinfo{journal}{Phys. Fluids} \textbf{\bibinfo{volume}{25}},
  \bibinfo{pages}{502} (\bibinfo{year}{1982}).

\bibitem[{\citenamefont{Villard et~al.}(2004)\citenamefont{Villard, Angelino,
  Bottino, Allfrey, Hatzky, Idomura, Sauter, and Tran}}]{villardFE04}
\bibinfo{author}{\bibfnamefont{L.}~\bibnamefont{Villard}},
  \bibinfo{author}{\bibfnamefont{P.}~\bibnamefont{Angelino}},
  \bibinfo{author}{\bibfnamefont{A.}~\bibnamefont{Bottino}},
  \bibinfo{author}{\bibfnamefont{S.~J.} \bibnamefont{Allfrey}},
  \bibinfo{author}{\bibfnamefont{R.}~\bibnamefont{Hatzky}},
  \bibinfo{author}{\bibfnamefont{Y.}~\bibnamefont{Idomura}},
  \bibinfo{author}{\bibfnamefont{O.}~\bibnamefont{Sauter}}, \bibnamefont{and}
  \bibinfo{author}{\bibfnamefont{T.~M.} \bibnamefont{Tran}},
  \bibinfo{journal}{Fusion Energy} \textbf{\bibinfo{volume}{46}},
  \bibinfo{pages}{B51} (\bibinfo{year}{2004}).

\bibitem[{\citenamefont{Candy et~al.}(2006)\citenamefont{Candy, Waltz, Parker,
  and Chen}}]{candy2PoP06}
\bibinfo{author}{\bibfnamefont{J.}~\bibnamefont{Candy}},
  \bibinfo{author}{\bibfnamefont{R.}~\bibnamefont{Waltz}},
  \bibinfo{author}{\bibfnamefont{S.~E.} \bibnamefont{Parker}},
  \bibnamefont{and} \bibinfo{author}{\bibfnamefont{Y.}~\bibnamefont{Chen}},
  \bibinfo{journal}{Phys. Plasmas} \textbf{\bibinfo{volume}{13}},
  \bibinfo{pages}{074501} (\bibinfo{year}{2006}).

\bibitem[{\citenamefont{Hinton and Hazeltine}(1976)}]{hintonRMP76}
\bibinfo{author}{\bibfnamefont{F.~L.} \bibnamefont{Hinton}} \bibnamefont{and}
  \bibinfo{author}{\bibfnamefont{R.~D.} \bibnamefont{Hazeltine}},
  \bibinfo{journal}{Rev. Mod. Phys.} \textbf{\bibinfo{volume}{25}},
  \bibinfo{pages}{239} (\bibinfo{year}{1976}).

\bibitem[{\citenamefont{Sugama et~al.}(1996)\citenamefont{Sugama, Okamoto,
  Horton, and Wakatani}}]{sugamaPoP96}
\bibinfo{author}{\bibfnamefont{H.}~\bibnamefont{Sugama}},
  \bibinfo{author}{\bibfnamefont{M.}~\bibnamefont{Okamoto}},
  \bibinfo{author}{\bibfnamefont{W.}~\bibnamefont{Horton}}, \bibnamefont{and}
  \bibinfo{author}{\bibfnamefont{M.}~\bibnamefont{Wakatani}},
  \bibinfo{journal}{Phys. Plasmas} \textbf{\bibinfo{volume}{3}},
  \bibinfo{pages}{2379} (\bibinfo{year}{1996}).

\bibitem[{\citenamefont{Howes et~al.}(2006)\citenamefont{Howes, Cowley,
  Dorland, Hammett, Quataert, and Schekochihin}}]{howesApJ06}
\bibinfo{author}{\bibfnamefont{G.~G.} \bibnamefont{Howes}},
  \bibinfo{author}{\bibfnamefont{S.~C.} \bibnamefont{Cowley}},
  \bibinfo{author}{\bibfnamefont{W.}~\bibnamefont{Dorland}},
  \bibinfo{author}{\bibfnamefont{G.~W.} \bibnamefont{Hammett}},
  \bibinfo{author}{\bibfnamefont{E.}~\bibnamefont{Quataert}}, \bibnamefont{and}
  \bibinfo{author}{\bibfnamefont{A.~A.} \bibnamefont{Schekochihin}},
  \bibinfo{journal}{Astrophys. J.} \textbf{\bibinfo{volume}{651}},
  \bibinfo{pages}{590} (\bibinfo{year}{2006}).

\bibitem[{\citenamefont{Jenko et~al.}(2000)\citenamefont{Jenko, Dorland,
  Kotschenreuther, and Rogers}}]{jenkoPoP00}
\bibinfo{author}{\bibfnamefont{F.}~\bibnamefont{Jenko}},
  \bibinfo{author}{\bibfnamefont{W.}~\bibnamefont{Dorland}},
  \bibinfo{author}{\bibfnamefont{M.}~\bibnamefont{Kotschenreuther}},
  \bibnamefont{and} \bibinfo{author}{\bibfnamefont{B.~N.}
  \bibnamefont{Rogers}}, \bibinfo{journal}{Phys. Plasmas}
  \textbf{\bibinfo{volume}{7}}, \bibinfo{pages}{1904} (\bibinfo{year}{2000}).

\bibitem[{\citenamefont{Candy and Waltz}(2006)}]{candyPoP06}
\bibinfo{author}{\bibfnamefont{J.}~\bibnamefont{Candy}} \bibnamefont{and}
  \bibinfo{author}{\bibfnamefont{R.}~\bibnamefont{Waltz}},
  \bibinfo{journal}{Phys. Plasmas} \textbf{\bibinfo{volume}{13}},
  \bibinfo{pages}{032310} (\bibinfo{year}{2006}).

\bibitem[{\citenamefont{Chen and Parker}(2007)}]{chenPoP07}
\bibinfo{author}{\bibfnamefont{Y.}~\bibnamefont{Chen}} \bibnamefont{and}
  \bibinfo{author}{\bibfnamefont{S.~E.} \bibnamefont{Parker}},
  \bibinfo{journal}{Phys. Plasmas} \textbf{\bibinfo{volume}{14}},
  \bibinfo{pages}{082301} (\bibinfo{year}{2007}).

\bibitem[{\citenamefont{Navkal et~al.}(2006)\citenamefont{Navkal, Ernst, and
  Dorland}}]{navkalAPS06}
\bibinfo{author}{\bibfnamefont{V.}~\bibnamefont{Navkal}},
  \bibinfo{author}{\bibfnamefont{D.~R.} \bibnamefont{Ernst}}, \bibnamefont{and}
  \bibinfo{author}{\bibfnamefont{W.}~\bibnamefont{Dorland}},
  \bibinfo{journal}{Bull. Am. Phys. Soc.}  (\bibinfo{year}{2006}),
  \bibinfo{note}{presented at the 48th Annual Meeting of the Division of Plasma
  Physics, October 30 - November 3, 2006, Philadelphia, PA, USA, JP1.00017}.

\bibitem[{\citenamefont{Rewoldt et~al.}(1986)\citenamefont{Rewoldt, Tang, and
  Hastie}}]{rewoldtPoF86}
\bibinfo{author}{\bibfnamefont{G.}~\bibnamefont{Rewoldt}},
  \bibinfo{author}{\bibfnamefont{W.~M.} \bibnamefont{Tang}}, \bibnamefont{and}
  \bibinfo{author}{\bibfnamefont{R.~J.} \bibnamefont{Hastie}},
  \bibinfo{journal}{Phys. Fluids} \textbf{\bibinfo{volume}{29}},
  \bibinfo{pages}{2893} (\bibinfo{year}{1986}).

\bibitem[{\citenamefont{Rutherford et~al.}(1970)\citenamefont{Rutherford,
  Kovrizhnikh, Rosenbluth, and Hinton}}]{rutherfordPRL70}
\bibinfo{author}{\bibfnamefont{P.~H.} \bibnamefont{Rutherford}},
  \bibinfo{author}{\bibfnamefont{L.~M.} \bibnamefont{Kovrizhnikh}},
  \bibinfo{author}{\bibfnamefont{M.~N.} \bibnamefont{Rosenbluth}},
  \bibnamefont{and} \bibinfo{author}{\bibfnamefont{F.~L.}
  \bibnamefont{Hinton}}, \bibinfo{journal}{Phys. Rev. Lett.}
  \textbf{\bibinfo{volume}{25}}, \bibinfo{pages}{1090} (\bibinfo{year}{1970}).

\bibitem[{\citenamefont{Catto and Tsang}(1976)}]{cattoPoF76}
\bibinfo{author}{\bibfnamefont{P.~J.} \bibnamefont{Catto}} \bibnamefont{and}
  \bibinfo{author}{\bibfnamefont{K.~T.} \bibnamefont{Tsang}},
  \bibinfo{journal}{Phys. Fluids} \textbf{\bibinfo{volume}{20}},
  \bibinfo{pages}{396} (\bibinfo{year}{1976}).

\bibitem[{\citenamefont{Abel et~al.}(2008)\citenamefont{Abel, Barnes, Cowley,
  Dorland, Hammett, and Schekochihin}}]{abelPoP08}
\bibinfo{author}{\bibfnamefont{I.~G.} \bibnamefont{Abel}},
  \bibinfo{author}{\bibfnamefont{M.}~\bibnamefont{Barnes}},
  \bibinfo{author}{\bibfnamefont{S.~C.} \bibnamefont{Cowley}},
  \bibinfo{author}{\bibfnamefont{W.}~\bibnamefont{Dorland}},
  \bibinfo{author}{\bibfnamefont{G.~W.} \bibnamefont{Hammett}},
  \bibnamefont{and} \bibinfo{author}{\bibfnamefont{A.~A.}
  \bibnamefont{Schekochihin}}, \bibinfo{journal}{Phys. Plasmas, submitted}
  (\bibinfo{year}{2008}), \bibinfo{note}{arXiv: 0806.1069}.

\bibitem[{\citenamefont{Hirshman and Sigmar}(1976)}]{hirshmanPoF76}
\bibinfo{author}{\bibfnamefont{S.}~\bibnamefont{Hirshman}} \bibnamefont{and}
  \bibinfo{author}{\bibfnamefont{D.}~\bibnamefont{Sigmar}},
  \bibinfo{journal}{Phys. Fluids} \textbf{\bibinfo{volume}{19}},
  \bibinfo{pages}{1532} (\bibinfo{year}{1976}).

\bibitem[{\citenamefont{Godunov}(1959)}]{GodunovMS59}
\bibinfo{author}{\bibfnamefont{S.~K.} \bibnamefont{Godunov}},
  \bibinfo{journal}{Mat. Sbornik} \textbf{\bibinfo{volume}{47}},
  \bibinfo{pages}{271} (\bibinfo{year}{1959}).

\bibitem[{\citenamefont{Richtmyer and Morton}(1967)}]{richtmyer}
\bibinfo{author}{\bibfnamefont{R.~D.} \bibnamefont{Richtmyer}}
  \bibnamefont{and} \bibinfo{author}{\bibfnamefont{K.~W.}
  \bibnamefont{Morton}}, \emph{\bibinfo{title}{Difference methods \\
  for initial
  value problems}} (\bibinfo{publisher}{Interscience}, \bibinfo{year}{1967}).

\bibitem[{\citenamefont{Sherman and Morrison}(1949)}]{shermanAMS49}
\bibinfo{author}{\bibfnamefont{J.}~\bibnamefont{Sherman}} \bibnamefont{and}
  \bibinfo{author}{\bibfnamefont{W.~J.} \bibnamefont{Morrison}},
  \bibinfo{journal}{Ann. Math. Stat.} \textbf{\bibinfo{volume}{20}},
  \bibinfo{pages}{621} (\bibinfo{year}{1949}).

\bibitem[{\citenamefont{Sherman and Morrison}(1950)}]{shermanAMS50}
\bibinfo{author}{\bibfnamefont{J.}~\bibnamefont{Sherman}} \bibnamefont{and}
  \bibinfo{author}{\bibfnamefont{W.~J.} \bibnamefont{Morrison}},
  \bibinfo{journal}{Ann. Math. Stat.} \textbf{\bibinfo{volume}{21}},
  \bibinfo{pages}{124} (\bibinfo{year}{1950}).

\bibitem[{\citenamefont{Degond and Lucquin-Desreux}(1994)}]{degondNM94}
\bibinfo{author}{\bibfnamefont{P.}~\bibnamefont{Degond}} \bibnamefont{and}
  \bibinfo{author}{\bibfnamefont{B.}~\bibnamefont{Lucquin-Desreux}},
  \bibinfo{journal}{Numer. Math.} \textbf{\bibinfo{volume}{68}},
  \bibinfo{pages}{239} (\bibinfo{year}{1994}).

\bibitem[{\citenamefont{Candy and Waltz}(2003)}]{candyJCP03}
\bibinfo{author}{\bibfnamefont{J.}~\bibnamefont{Candy}} \bibnamefont{and}
  \bibinfo{author}{\bibfnamefont{R.~E.} \bibnamefont{Waltz}},
  \bibinfo{journal}{J. Comp. Phys.} \textbf{\bibinfo{volume}{186}},
  \bibinfo{pages}{545} (\bibinfo{year}{2003}).

\bibitem[{\citenamefont{Durran}(1999)}]{durran}
\bibinfo{author}{\bibfnamefont{D.~R.} \bibnamefont{Durran}},
  \emph{\bibinfo{title}{Numerical methods for wave equations in \\
  geophysical
  fluid dynamics}} (\bibinfo{publisher}{Springer}, \bibinfo{year}{1999}).

\bibitem[{\citenamefont{Hildebrand}(1987)}]{hilde}
\bibinfo{author}{\bibfnamefont{F.~B.} \bibnamefont{Hildebrand}},
  \emph{\bibinfo{title}{Introduction to Numerical Analysis}}
  (\bibinfo{publisher}{Dover}, \bibinfo{year}{1987}).

\bibitem[{\citenamefont{Helander and Sigmar}(2002)}]{helander}
\bibinfo{author}{\bibfnamefont{P.}~\bibnamefont{Helander}} \bibnamefont{and}
  \bibinfo{author}{\bibfnamefont{D.~J.} \bibnamefont{Sigmar}},
  \emph{\bibinfo{title}{Collisional transport in \\
  magnetized plasmas}}
  (\bibinfo{publisher}{Cambridge University Press}, \bibinfo{year}{2002}).

\bibitem[{\citenamefont{Barnes}(1966)}]{barnesPoF66}
\bibinfo{author}{\bibfnamefont{A.}~\bibnamefont{Barnes}},
  \bibinfo{journal}{Phys. Fluids} \textbf{\bibinfo{volume}{9}},
  \bibinfo{pages}{1483} (\bibinfo{year}{1966}).

\bibitem[{\citenamefont{Freidberg}(1987)}]{freidbergMHD}
\bibinfo{author}{\bibfnamefont{J.~P.} \bibnamefont{Freidberg}},
  \emph{\bibinfo{title}{Ideal Magnetohydrodynamics}}
  (\bibinfo{publisher}{Plenum}, \bibinfo{year}{1987}).

\bibitem[{\citenamefont{Kadomtsev}(1960)}]{kadomtsevSP60}
\bibinfo{author}{\bibfnamefont{B.~B.} \bibnamefont{Kadomtsev}},
  \bibinfo{journal}{Sov. Phys. JETP} \textbf{\bibinfo{volume}{10}},
  \bibinfo{pages}{780} (\bibinfo{year}{1960}).

\bibitem[{\citenamefont{Kesner}(2000)}]{kesnerPoP00}
\bibinfo{author}{\bibfnamefont{J.}~\bibnamefont{Kesner}},
  \bibinfo{journal}{Phys. Plasmas} \textbf{\bibinfo{volume}{7}},
  \bibinfo{pages}{3837} (\bibinfo{year}{2000}).

\bibitem[{\citenamefont{Simakov et~al.}(2001)\citenamefont{Simakov, Catto, and
  Hastie}}]{simakovPoP01}
\bibinfo{author}{\bibfnamefont{A.~N.} \bibnamefont{Simakov}},
  \bibinfo{author}{\bibfnamefont{P.~J.} \bibnamefont{Catto}}, \bibnamefont{and}
  \bibinfo{author}{\bibfnamefont{R.~J.} \bibnamefont{Hastie}},
  \bibinfo{journal}{Phys. Plasmas} \textbf{\bibinfo{volume}{8}},
  \bibinfo{pages}{4414} (\bibinfo{year}{2001}).

\bibitem[{\citenamefont{Simakov et~al.}(2002)\citenamefont{Simakov, Hastie, and
  Catto}}]{simakovPoP02}
\bibinfo{author}{\bibfnamefont{A.~N.} \bibnamefont{Simakov}},
  \bibinfo{author}{\bibfnamefont{R.~J.} \bibnamefont{Hastie}},
  \bibnamefont{and} \bibinfo{author}{\bibfnamefont{P.~J.} \bibnamefont{Catto}},
  \bibinfo{journal}{Phys. Plasmas} \textbf{\bibinfo{volume}{9}},
  \bibinfo{pages}{201} (\bibinfo{year}{2002}).

\bibitem[{\citenamefont{Kesner and Hastie}(2002)}]{kesnerPoP02}
\bibinfo{author}{\bibfnamefont{J.}~\bibnamefont{Kesner}} \bibnamefont{and}
  \bibinfo{author}{\bibfnamefont{R.~J.} \bibnamefont{Hastie}},
  \bibinfo{journal}{Phys. Plasmas} \textbf{\bibinfo{volume}{9}},
  \bibinfo{pages}{395} (\bibinfo{year}{2002}).

\bibitem[{\citenamefont{Ricci et~al.}(2006{\natexlab{a}})\citenamefont{Ricci,
  Rogers, Dorland, and Barnes}}]{ricciPoP06}
\bibinfo{author}{\bibfnamefont{P.}~\bibnamefont{Ricci}},
  \bibinfo{author}{\bibfnamefont{B.~N.} \bibnamefont{Rogers}},
  \bibinfo{author}{\bibfnamefont{W.}~\bibnamefont{Dorland}}, \bibnamefont{and}
  \bibinfo{author}{\bibfnamefont{M.}~\bibnamefont{Barnes}},
  \bibinfo{journal}{Phys. Plasmas} \textbf{\bibinfo{volume}{13}},
  \bibinfo{pages}{062102} (\bibinfo{year}{2006}{\natexlab{a}}).

\bibitem[{\citenamefont{Ricci et~al.}(2006{\natexlab{b}})\citenamefont{Ricci,
  Rogers, and Dorland}}]{ricciPRL06}
\bibinfo{author}{\bibfnamefont{P.}~\bibnamefont{Ricci}},
  \bibinfo{author}{\bibfnamefont{B.~N.} \bibnamefont{Rogers}},
  \bibnamefont{and} \bibinfo{author}{\bibfnamefont{W.}~\bibnamefont{Dorland}},
  \bibinfo{journal}{Phys. Rev. Lett.} \textbf{\bibinfo{volume}{97}},
  \bibinfo{pages}{245001} (\bibinfo{year}{2006}{\natexlab{b}}).

\bibitem[{\citenamefont{Dimits et~al.}(2000)\citenamefont{Dimits, Bateman,
  Beer, Cohen, Dorland, Hammett, et~al.}}]{dimitsPoP00}
\bibinfo{author}{\bibfnamefont{A.~M.} \bibnamefont{Dimits}},
  \bibinfo{author}{\bibfnamefont{G.}~\bibnamefont{Bateman}},
  \bibinfo{author}{\bibfnamefont{M.~A.} \bibnamefont{Beer}},
  \bibnamefont{et~al.},
  \bibinfo{journal}{Phys. Plasmas} \textbf{\bibinfo{volume}{7}},
  \bibinfo{pages}{969} (\bibinfo{year}{2000}).

\bibitem[{\citenamefont{Tatsuno and Dorland}(2008)}]{tatsunoAN08}
\bibinfo{author}{\bibfnamefont{T.}~\bibnamefont{Tatsuno}} \bibnamefont{and}
  \bibinfo{author}{\bibfnamefont{W.}~\bibnamefont{Dorland}},
  \bibinfo{journal}{Astron. Nachr.} \textbf{\bibinfo{volume}{329}},
  \bibinfo{pages}{688} (\bibinfo{year}{2008}).

\end{thebibliography}
\end{document}